\documentclass[final,5p,times,authoryear]{elsarticle}
\usepackage{framed,multirow}

\usepackage{amssymb}
\usepackage{latexsym}
\usepackage{booktabs}
\usepackage{tikz}
\usepackage{amsmath}
\usepackage{float}
\usepackage{csquotes}

\usepackage[labelfont=md,textfont=md,font=normalsize]{caption}

\usepackage{url}
\usepackage[breaklinks]{hyperref}

\journal{Medical Image Analysis}

\begin{document}
\begin{frontmatter}
\title{Cross-modality image synthesis from TOF-MRA to CTA using diffusion-based models}

\author[1]{Alexander Koch\corref{cor1}}
\cortext[cor1]{Corresponding author}
\ead{alexander.koch@charite.de}

\author[1]{Orhun Utku Aydin}
\author[1]{Adam Hilbert}
\author[1]{Jana Rieger}
\author[1,2]{Satoru Tanioka}
\author[3]{Fujimaro Ishida}
\author[1,4]{Dietmar Frey}

\affiliation[1]{
    organization={CLAIM - Charité Lab for AI in Medicine, Charité – Universitätsmedizin Berlin, corporate member of Freie Universität Berlin and Humboldt-Universität zu Berlin},
    addressline={Charitéplatz 1},
    postcode={101117},
    city={Berlin},
    country={Germany}}

\affiliation[2]{
    organization={Department of Neurosurgery, Mie University Graduate School of Medicine},
    addressline={2-174 Edobashi},
    postcode={514-8507},
    city={Tsu},
    country={Japan}}

\affiliation[3]{
    organization={Department of Neurosurgery, Mie Chuo Medical Center},
    addressline={2158-5 Myojin-cho},
    postcode={514-1101},
    city={Hisai, Tsu},
    country={Japan}}

\affiliation[4]{
    organization={Department of Neurosurgery, Charité – Universitätsmedizin Berlin, corporate member of Freie Universität Berlin and Humboldt-Universität zu Berlin},
    addressline={Charitéplatz 1},
    postcode={101117},
    city={Berlin},
    country={Germany}}

\begin{abstract}
Cerebrovascular disease often requires multiple imaging modalities for accurate diagnosis, treatment, and monitoring. Computed Tomography Angiography (CTA) and Time-of-Flight Magnetic Resonance Angiography (TOF-MRA) are two common non-invasive angiography techniques, each with distinct strengths in accessibility, safety, and diagnostic accuracy. While CTA is more widely used in acute stroke due to its faster acquisition times and higher diagnostic accuracy, TOF-MRA is preferred for its safety, as it avoids radiation exposure and contrast agent-related health risks. Despite the predominant role of CTA in clinical workflows, there is a scarcity of open-source CTA data, limiting the research and development of AI models for tasks such as large vessel occlusion detection and aneurysm segmentation. This study explores diffusion-based image-to-image translation models to generate synthetic CTA images from TOF-MRA input. We demonstrate the modality conversion from TOF-MRA to CTA and show that diffusion models outperform a traditional U-Net-based approach. Our work compares different state-of-the-art diffusion architectures and samplers, offering recommendations for optimal model performance in this cross-modality translation task.
\end{abstract}

\end{frontmatter}

\section{Introduction}

Vessel neuroimaging techniques provide crucial information for diagnosis, treatment and monitoring of cerebrovascular disease. The two main noninvasive angiography modalities, Computed Tomography Angiography (CTA) and Time-of-Flight Magnetic Resonance Angiography (TOF-MRA) have unique advantages and disadvantages in terms of acquisition time, accessibility, safety and diagnostic accuracy \citep{demchuk_comparing_2016}. For example, whereas CTA offers better accessibility, faster acquisition times and higher diagnostic accuracy for collateral assessment, TOF-MRA provides greater safety as it avoids radiation exposure, and potential side effects of contrast agents such as allergic reactions or contrast-induced nephropathy. 

Deep learning has increasingly been used to analyse angiography images in acute stroke with commercial tools available for large vessel occlusion detection, and collateral score assessment \citep{soun_artificial_2021}. As CTA is the predominant imaging modality for acute stroke, nearly all commercially available tools are developed for CTA workflows. However, this stands in stark contrast to the limited availability of open-source CTA data compared to the TOF-MRA modality \citep{yang_benchmarking_2024} for research purposes. Data scarcity concerning CTA significantly hampers the development of AI models targeting the CTA modality such as vessel and aneurysm segmentation tools. 

Image-to-image translation offers a promising option to generate a synthetic target modality from an available input modality. Different architectures have been proposed for both paired and unpaired image-to-image translation tasks. Unpaired image-to-image translation is based on transferring the style between either images of different subjects or unaligned/unregistered images of the same subject. Paired image-to-image translation on the other hand aims to find a direct mapping between two aligned images of the same subject. Generative adversarial networks (GANs) and diffusion-based models constitute the current state of the art for paired image-to-image translation tasks \citep{saharia_palette_2022,zhou_cascaded_2024}. Recently, diffusion models have been adopted due to their improved training and arguably higher image quality compared to GANs \citep{kazerouni_diffusion_2023}. However, they also come with their unique challenges.

Applications of image-to-image translation in medical imaging aim to solve a wide range of clinical problems. For instance, they aim to reduce radiation exposure \citep{zhou_limited_2021}, enhance images with virtual contrast agents \citep{rofena_deep_2024}, and increase generalization of segmentation models \citep{sandfort_dataaugcyclegan_2019}. Prior works have addressed various intra-modality (DWI to FLAIR) \citep{benzakoun_synthetic_2022}, and cross-modality (CT to MRI) conversion tasks \citep{liu_multicyclegan_2021}. However, to the best of our knowledge, no prior work has explored the inter-modality translation task of synthesizing CTA images from TOF-MRA input. Therefore, to address this research gap, we set out to explore diffusion-based image-to-image translation models from TOF-MRA to CTA. In this work we:
\begin{enumerate}
    \item show that paired TOF-MRA to CTA modality conversion is feasible using deep learning on 2D slices
    \item demonstrate that diffusion models outperform standard dense-prediction/ U-Net-based approaches
    \item compare different state-of-the-art diffusion architectures and samplers and provide recommendations for optimal results
\end{enumerate}

\section{Background}

\subsection{Existing work on cross-modality image synthesis}

Most work in cross-modality image synthesis relies on GAN-based methods and operates on MRI data \citep{xie24xmodalitysurvey}. \citet{maspero18pix2pix} use a pix2pix model to generate CT out of MRI imaging. \citet{olut18mra} generate MRA imaging out of T1 and T2 imaging, employing the pix2pix model. Further, \citet{zhang18cyclegan} use a CycleGAN model to perform 3D cross-modality image synthesis on CT and MRI imaging. Recently diffusion models have become more widely used due to higher quality image synthesis compared to GANs. \citet{zhu23makeavolume} use a mixed 2D and 3D approach using a latent diffusion model, converting between SWI and MRA. Further they evaluate their model to perform synthesis between T1 and T2 volumes. \citet{lyu22ctmri} show that diffusion models can compete with CNN and GAN-based methods and generate between MRI and CT. Moreover, they also use different sampling methods such as Euler-Maruyama, Predictor-Corrector and using the explicit Runge-Kutta method. \citet{zhou_cascaded_2024} propose a combination of a GAN and a diffusion model for high-quality medical image-to-image translation.

\subsection{Diffusion Models}

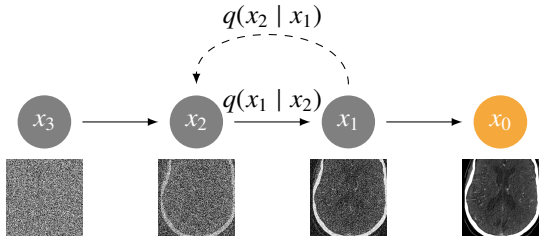
\begin{figure}[h]
    \centering
    \begin{tikzpicture}
    \definecolor{color1}{HTML}{f5a93d}
    \definecolor{color2}{HTML}{5e8dfd}

    \node[circle, fill=gray, radius=20] (x3) at (1,4) {\textcolor{white}{$x_3$}};
    \node[circle, fill=gray, radius=20] (x2) at (3,4) {\textcolor{white}{$x_2$}};
    \node[circle, fill=gray, radius=20] (x1) at (5,4) {\textcolor{white}{$x_1$}};
    \node[circle, fill=color1, radius=20] (x0) at (7,4) {\textcolor{white}{$x_0$}};

    \node (img3) at (1,3) {\includegraphics[width=1cm, angle=-90]{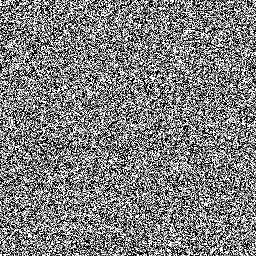}};
    \node (img2) at (3,3) {\includegraphics[width=1cm, angle=-90]{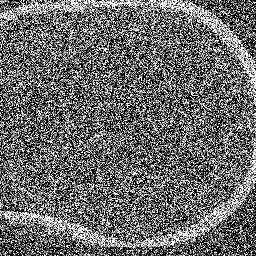}};
    \node (img1) at (5,3) {\includegraphics[width=1cm, angle=-90]{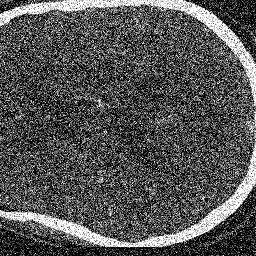}};
    \node (img0) at (7,3) {\includegraphics[width=1cm, angle=-90]{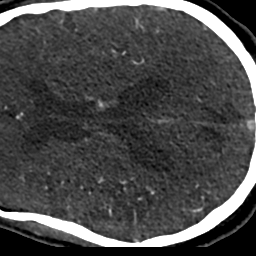}};

    \draw[->,>=latex,shorten >= 4pt, shorten <= 4pt] (x3) -- (x2);
    \draw[->,>=latex,shorten >= 4pt, shorten <= 4pt] (x2) -- node[above] {$q(x_1 \mid x_2)$} (x1);
    \draw[->,>=latex,shorten >= 4pt, shorten <= 4pt] (x1) -- (x0);
    \draw[->, >=latex, bend right=90, distance=1cm, dashed,shorten >= 4pt, shorten <= 4pt] (x1.north) to node[above] {$q(x_2 \mid x_1)$} (x2.north);
    \end{tikzpicture}
    
    \caption{\textbf{Diffusion process.} Graphical model for the (Markovian) diffusion process. The forward process going from $x_0$ to $x_3$ progressively adds noise. The backwards (denoising) process going from $x_3$ to $x_0$ successively removes noise.}
    \label{fig:denoising}
\end{figure}

A diffusion process is a Markov chain, which adds Gaussian noise over time. We follow the definition used by \citet{hoogeboom_simple_2023}. The forward process is described as follows:
\begin{align} \label{eqn_marginal}
    q(z_t \mid x) &= \mathcal{N}(z_t \mid \alpha_t x, \sigma_t^2 I)
\end{align}

where $\alpha_t, \sigma_t \in (0,1)$ are hyperparameters for the noise schedule. The parameter $\alpha_t$ is increasing over time, while $\sigma_t$ is decreasing. We are using a variance preserving processes \citep{song21scorebased}, i.e\ $\alpha_t^2 = 1 - \sigma_t^2$. The forward transition distribution is given by
\begin{align} \label{eqn_transition}
    q(z_t \mid z_s) &= \mathcal{N}(z_t \mid \alpha_{ts} z_s, \sigma_{ts}^2 I)
\end{align}

where $\alpha_{ts} = \alpha_t / \alpha_s$ and $\sigma_{ts}^2 = \sigma_t^2 - \alpha_{ts}^2 \sigma_s^2$ and $t > s$.

A common noise schedule is the cosine schedule \citep{nichol21improved} defined as $\alpha_t = \cos(\pi t / 2)$, which under the assumption of a variance preserving process implies $\sigma_t = \sin(\pi t / 2)$. The \textit{signal-to-noise-ratio} (SNR) is given by $\mathrm{SNR}(t) = \alpha_t^2 / \sigma_t^2 = \tan(\pi t / 2)^{-2}$. Thus, in log-space we can write $\log \mathrm{SNR}(t) = -2 \log \tan(\pi t / 2)$ and the hyperparameters are given by $\alpha_t^2=\mathrm{sigmoid}(\log \mathrm{SNR}(t))$ and $\sigma_t^2=\mathrm{sigmoid}(-\log \mathrm{SNR}(t))$.

The denoising process is defined as follows:
\begin{align} \label{eqn_denoising}
    q(z_s \mid z_t, x) = \mathcal{N}(\mu_{t \rightarrow s}, \sigma_{t \rightarrow s}^2 I)
\end{align}

where $\mu_{t \rightarrow s} = \frac{\alpha_{ts} \sigma_s^2}{\sigma_t^2} z_t + \frac{\alpha_s \sigma_{ts}^2}{\sigma_t^2} x$ and $\sigma_{t \rightarrow s}^2 = \frac{\sigma_{ts}^2 \sigma_s^2}{\sigma_t^2}$.

To train a model one can choose multiple prediction targets. One can choose to predict the start $x$ by the approximation $\hat{x} = f_\theta(z_t)$. Further one can choose to predict the noise, relying on equation \ref{eqn_marginal} by using the re-parameterization trick $z_t = \alpha_t x + \sigma_t \epsilon_t$, where $\epsilon_t \sim \mathcal{N}(0, I)$. To obtain the start from the prediction $\hat{\epsilon_t} = f_\theta(z_t)$, one can then use $\hat{x} = (z_t - \sigma_t \hat{\epsilon_t}) / \alpha_t$. Another approach which proved to be more stable is to predict the velocity v, also known as \textit{v prediction}, introduced by \citet{salimans_progressive_2022}, defined as $v_t = \alpha_t \epsilon_t - \sigma_t x$. The model then predicts $\hat{v_t} = f_\theta(z_t)$, which gives $\hat{x} = \alpha_t z_t - \sigma_t \hat{v_t}$. To train a denoising model one then minimizes
\begin{align} \label{eqn_loss}
\mathbb{E}_{\epsilon,t} [ w(t) \lVert \hat{x}_\theta(z_t) - x \rVert_2^2 ]
\end{align}

which directly optimizes the prediction of the start. \citet{ho_denoising_2020} choose to predict and minimize epsilon directly, which gives
\begin{align} \label{eqn_eps}
L_\theta &= \lVert \hat{\epsilon}_\theta(z_t) - \epsilon \rVert_2^2 = \frac{\alpha^2}{\sigma^2} \lVert \hat{x}_\theta(z_t) - x \rVert_2^2
\end{align}
creating the weighting of $w(t) = \mathrm{SNR}(t)$. When predicting the velocity v, we get an implicit weighting function of $w(t) = 1 + \alpha^2 / \sigma^2 = 1 + \mathrm{SNR}(t)$. The choice of the weighting greatly impacts convergence and model performance \citep{hang_efficient_2023}.

\subsection{Samplers}

As an alternative to the previously described approach of using the equation \ref{eqn_denoising} for denoising, we can use different sampling methods, to further improve the results or to reduce the sampling time.
One such sampler is the Denoising Diffusion Implicit Model (DDIM), introduced by \citet{song21ddim}. This turns the existing model into an implicit probabilistic model. The generative process then becomes deterministic, except for the first step. Samples are deterministically generated from latent variables. The DDIM sampler can be seen as a linearization of the probability flow ordinary differential equation used in diffusion models \citep{salimans_progressive_2022,heek24}. The DDIM update rule is given by:
\begin{align} \label{eqn_ddim}
    z_s &= \alpha_s x + \frac{\sigma_s}{\sigma_t}(z_t - \alpha_t x)
\end{align}

In this paper, we refer to the method in \ref{eqn_denoising} as DDPM (Denoising Diffusion Probabilistic Model) sampling and \ref{eqn_ddim} as DDIM sampling. More complex ordinary differential equation solvers can be tried to obtain better results. In this paper we limit ourselves to DDPM and DDIM.

\section{Method}

\subsection{Data}

We use data from the Topology-Aware Anatomical Segmentation of the Circle of Willis (TopCoW) challenge \citep{yang_topcow_2023}. The dataset\footnote{TopCoW challenge data available from \url{https://topcow23.grand-challenge.org/data/}} comprises patients admitted to the Stroke Center of the University Hospital Zurich and provides paired CTA and TOF-MRA imaging. The available data is already anonymized and defaced. We co-register the TOF-MRA images non-linearly to the CTA images using ANTs \citep{tustison_antsx_2021} with B-spline interpolation. Overall we work with 89 patients and split them into 62 for training and 13 for validation and 14 for testing. We train all models as slice-wise two-dimensional models. We filter out slices that have low overlap between the source and target modality or have less than 200 pixels. In total, we use 10737 slices for training, 2162 for validation and 2319 for testing. For the CTA images we perform windowing such that the density is in range $[-50, 350]$. Both modalities, CTA and TOF-MRA images are min-max scaled to be in $[-1,1]$ range.

We validate our models on a private external test set. This test set consists of 11 patients with both TOF-MRA and CTA imaging. Ethical approval for the data usage was obtained from the institutional review board at Mie Chuo Medical Center (approval number: 2023-53). For detailed imaging parameters of the dataset see table \ref{tab:appendix_imaging_parameters}. Again, we co-register the TOF-MRA imaging data to the CTA imaging data.

\subsection{Architectures}

The first model architecture is a U-Net \citep{ronneberger_u-net_2015} for image-to-image translation which directly tries to predict the target modality. This is our baseline model. We use a single residual block per resolution and 128 base channels. Group normalization is applied with 32 groups.

Second, we use an Ablated diffusion model (ADM) \citep{dhariwal_diffusion_2021}. The architecture has 128 base channels, two residual blocks per resolution and multi-resolution self-attention with four attention heads.

Third, we compare the models to the U-ViT architecture proposed by \cite{hoogeboom_simple_2023}. Here, the middle part of the ADM is replaced by a transformer. Additionally, the image is first deconstructed using a discrete wavelet transform before passing it to the U-Net and later reconstructed using the same wavelet kernel to perform the final prediction. The transformer has 16 layers with 4 attention heads and a sinusoidal positional embedding. Gated Linear Units \citep{dauphin_language_2017} with Swish activation (SwiGLU) proposed in \cite{shazeer_glu_2020} are used as MLP blocks. For the discrete wavelet transform we use the Cohen-Daubechies-Feauveau (CDF) wavelet 9/7, also referred to as the biorthogonal 4, 4 wavelet. We apply one level of deconstruction.

Lastly, we train a Diffusion Transformer, introduced by \citet{peebles23dit}. This architecture uses a standard vision transformer and an adaptive layer normalization to scale the denoising timestep embedding. As with the U-ViT, we again employ a sinusoidal positional embedding and SwiGLUs for the MLPs. We use the DiT-L (Large) configuration with a patch size of 16, i.e.\ a hidden size of 1024, 16 attention heads and a depth of 24. The original paper applies this architecture only as a model operating on the latent space of a pre-trained autoencoder. However, the authors state that the method should also work on the pixel-level.

All models use pixel-shuffle downsampling and upsampling \citep{shi2016pixelshuffle}, instead of transposed convolutions or bilinear interpolations followed by convolutions. The U-Net, ADM and U-ViT each apply three stages of downsampling and upsampling respectively, with $[C, 2C, 4C]$ channels per stage, where $C=128$ is the number of channels. Moreover, all diffusion models use Root Mean Square Layer Normalization (RMSNorm) \citep{zhang_root_2019}. RMSNorm has been shown to be the best performing normalization variant for transformers \citep{narang_transformer_2021} and has recently been shown to work well for diffusion-based models too \citep{karras_analyzing_2024}.

\subsection{Diffusion Setup}

For all diffusion models, we use an $\alpha$-cosine noise schedule \citep{nichol21improved} and \textit{v prediction} parameterization, as it has been found that training using this parameterization is more reliable \citep{hoogeboom_simple_2023}. We directly minimize v as opposed to the standard epsilon loss. To accelerate convergence, we apply the Min-SNR loss weighting strategy \citep{hang_efficient_2023}. The weighting function is defined as
\begin{align} \label{eqn_minsnr_eps}
    w_\mathrm{v}(t) &= \frac{\min \{ \mathrm{SNR}(t), \gamma \}}{\mathrm{SNR(t)}+1}
\end{align}

where we set $\gamma = 5$ as used in the original paper. It forces the model to pay less attention to small noise levels. Notice that we divide by $\mathrm{SNR}(t)+1$, since we are minimizing v, removing the imposed implicit weighting.

The models are conditioned on the source modality by providing the image as a separate channel. Noise timesteps are embedded using a shared sinusoidal position embedding. We perform clipping after each diffusion step to $[-1,1]$ range to stabilize sampling and avoid divergent behaviour.

\subsection{Evaluation metrics}

To evaluate the model, we utilize common metrics in medical imaging which are used for cross-modality image synthesis. We evaluate the model using Peak Signal-to-Noise Ratio (PSNR), Scale Structural Similarity Index Measure (SSIM), mean squared error (MSE) and mean absolute error (MAE). To measure perceptual similarity we measure the Fréchet distance. Contrary to the common trend of using pre-trained networks based on medical features (such as MedicalNet or RadImageNet) we apply a network pre-trained on ImageNet. Recently, it has been found that ImageNet trained predictors are more reliable and align more with human judgement than feature extractors based on medical datasets \citep{woodland_feature_2024}. We use a ViT-B/16 \citep{dosovitskiy21vit} pre-trained on ImageNet-21k from the official google repository\footnote{Weights and model taken from \url{https://github.com/google-research/vision_transformer}} as our feature extractor. We remove the final layer and utilize the features of the class token. The FD score is calculated on the test set.

\subsection{Training}

We train all three architectures on $256 \times 256$ random crops of the original slices. All models are trained using the Adam optimizer \citep{kingma_adam2015} with a constant learning rate of $1 \times 10^{-4}$ and a batch size of 16. No augmentation, weight decay or other forms of regularization are used. Training is performed in bfloat16 precision, for a total budget of 150K steps. We implement all our models in Flax \citep{heek23flax} on top of JAX \citep{jax2018github} with Optax \citep{deepmind2020jax}. Our implementation and pre-trained models are available at \url{https://github.com/alexander-koch/xmodality}.

\subsection{Volume reconstruction}

Since we do not have a direct way to sample 3D CTA images from TOF-MRA, we have to perform the reconstruction in two dimensions. This assumes that the trained models provide high robustness and accuracy. There are multiple possible ways to perform inference with our models on full scans.

While we could simply iterate through each slice of the TOF-MRA image in full resolution and perform inference on our models, this is likely to produce bad results, since this requires the models to handle resolutions that they have not seen during training. Instead, we resample the image slice-wise to have a resolution of $256 \times 256$, performing inference on the model and afterwards rescaling the image to its original size. Despite having to downsample and potentially blur out the image, we find that this produces high-quality images. For downsampling and upsampling of the slices we use prefiltered cubic spline interpolation.

\section{Results}

\subsection{Evaluation of the models}

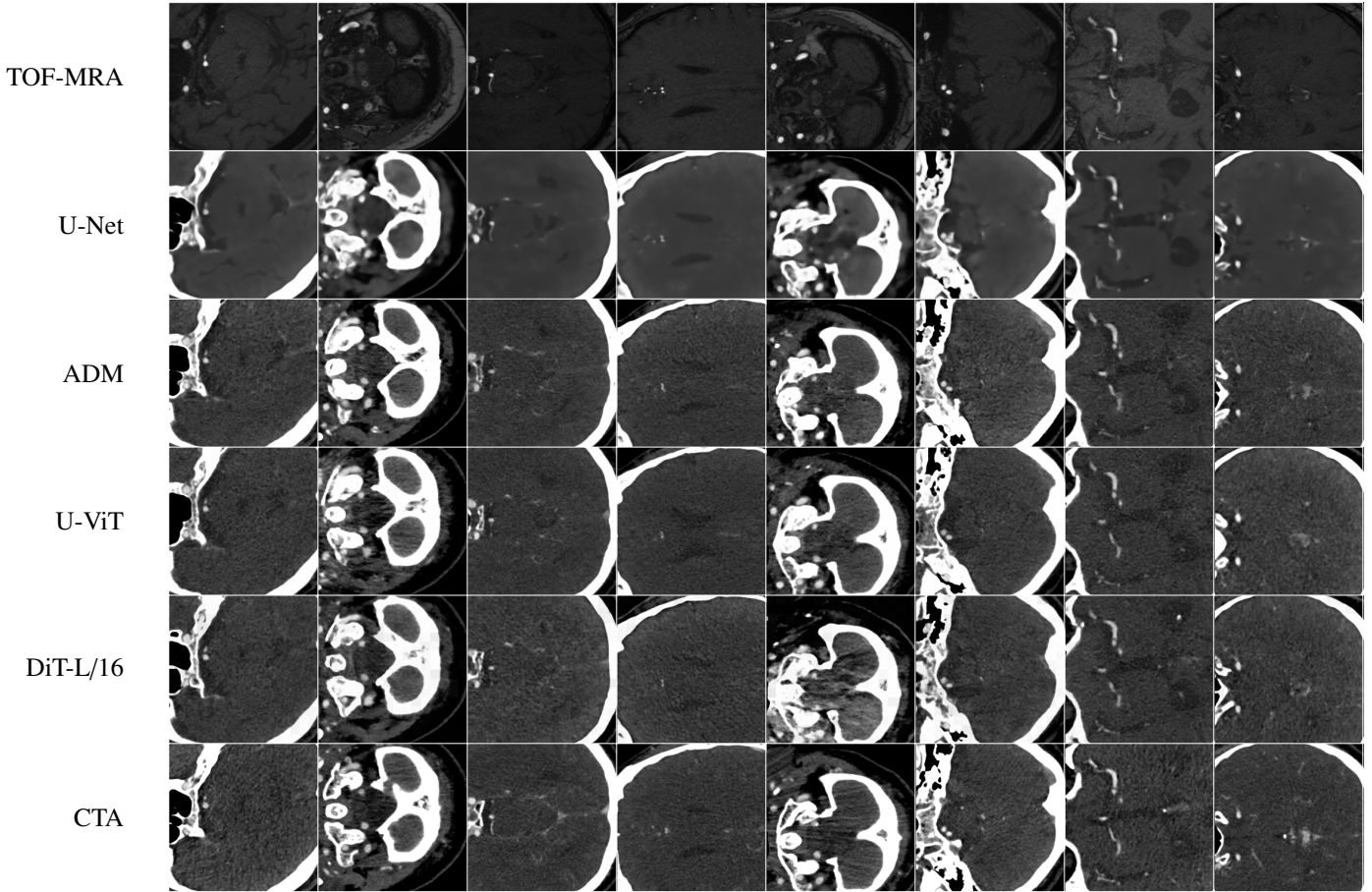
\begin{figure*}[h]
    \centering
    \begin{tikzpicture}
    \draw (0, 0) node[anchor=south west,inner sep=0] {\includegraphics[height=12cm, width=16cm]{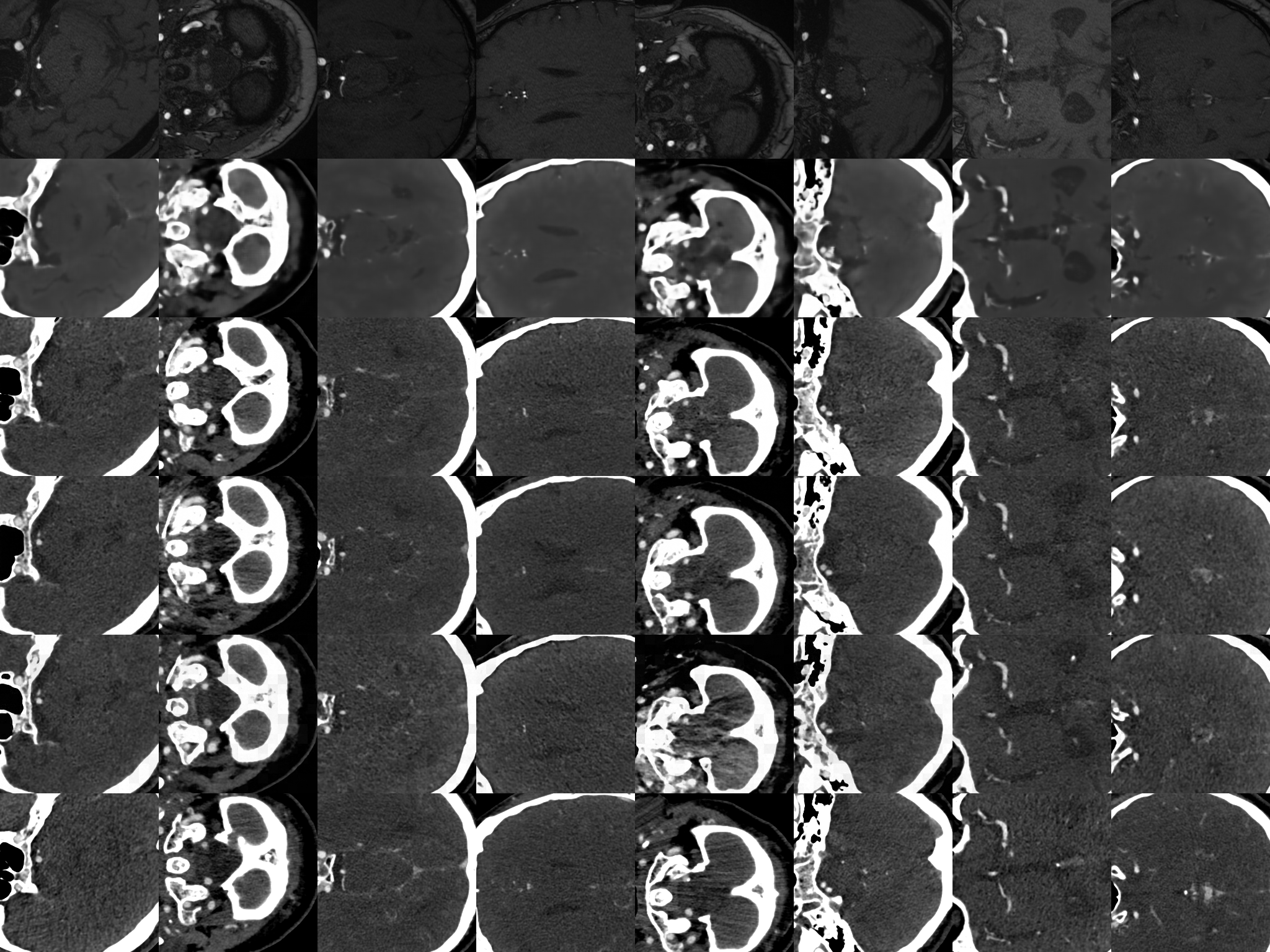}};
    \draw [help lines, step=2.0cm, color=white] (0,0) grid (16,12);
    \draw (-1.4,11) node[align=right, text width=1.6cm] {TOF-MRA};
    \draw (-1.4,9) node[align=right, text width=1.6cm] {U-Net};
    \draw (-1.4,7) node[align=right, text width=1.6cm] {ADM};
    \draw (-1.4,5) node[align=right, text width=1.6cm] {U-ViT};
    \draw (-1.4,3) node[align=right, text width=1.6cm] {DiT-L/16};
    \draw (-1.4,1) node[align=right, text width=1.6cm] {CTA};
    \end{tikzpicture}
    \caption{\textbf{Model outputs.} Comparison of each model using the same initial noise on the random samples of test set using the same random seed. Each sample is generated using 128 DDPM sampling steps. Each row represents one model output, except the first row and the last, which are the source and target image.}
    \label{fig:architecture_images}
\end{figure*}

\begin{table}[h]
\renewcommand{\arraystretch}{1.4}
\setlength\tabcolsep{4pt}

\caption{\textbf{Model results.} For each model we compute the average metrics on the test set. For the diffusion-based models, we use 1000 sampling steps. The metrics are computed on normalized images in $[0,1]$ range. We underline the second best results.}

\scalebox{0.85}{
\begin{tabular}{@{}lllllll@{}}
\toprule
Model & Sampler & MSE ($\downarrow$) & MAE ($\downarrow$) & PSNR ($\uparrow$) & SSIM ($\uparrow$) & FD ($\downarrow$) \\ \midrule
U-Net & N/A & \textbf{0.020} & \textbf{0.067} & \textbf{19.306} & \textbf{0.528} & 10.518 \\ \cmidrule{1-7}
ADM & DDPM & 0.027 & 0.075 & 17.949 & \underline{0.458} & 1.005 \\
U-ViT & DDPM & \underline{0.025} & \underline{0.073} & \underline{18.062} & 0.454 & 0.919 \\
DiT-L/16 & DDPM & 0.027 & 0.079 & 17.647 & 0.422 & \textbf{0.816} \\ \cmidrule{1-7}
ADM & DDIM & 0.029 & 0.079 & 17.573 & 0.440 & 0.947 \\
U-ViT & DDIM & 0.026 & 0.074 & 17.941 & 0.445 & \underline{0.910} \\
DiT-L/16 & DDIM & 0.029 & 0.082 & 17.321 & 0.398 & 1.904 \\
\bottomrule
\end{tabular}
}
\label{tab:model_results}
\end{table}

Since computing slices of variable sizes is very time-consuming, we fix one random seed and perform evaluation of the test set of the corresponding fixed random crops of $256 \times 256$.

In table \ref{tab:model_results} we see the metrics of the individual models. While the standard U-Net achieves slightly better performance in the intensity-based metrics, the diffusion models have a much lower FD score. The FD scores are $10.518$ for the U-Net, $1.005$ for the ADM, $0.919$ for the U-ViT and $0.816$ for the DiT-L/16. Using DDIM sampling we obtain FD scores of $0.947$ for the ADM, $0.91$ for the U-ViT and $1.904$ for the DiT-L/16. Visually, the U-Net creates washed out images that are structurally sound, but do not appear realistic texture-wise (see figure \ref{fig:architecture_images}).

The U-ViT has a better PSNR of $18.062$ compared to the ADM with PSNR of $17.949$, but a worse SSIM of $0.454$ instead of $0.458$ for the ADM. The MSE and MAE scores are $0.025$ and $0.073$ respectively, lower than both the DiT-L/16 and the ADM. The DiT-L/16 has the best FD score of $0.816$, while being slightly worse in every metric compared to the other diffusion models. It requires the most parameters, while being the fastest diffusion model in terms of speed (see table \ref{tab:model_compute}).

\begin{figure}[h]
    \centering
    \includegraphics{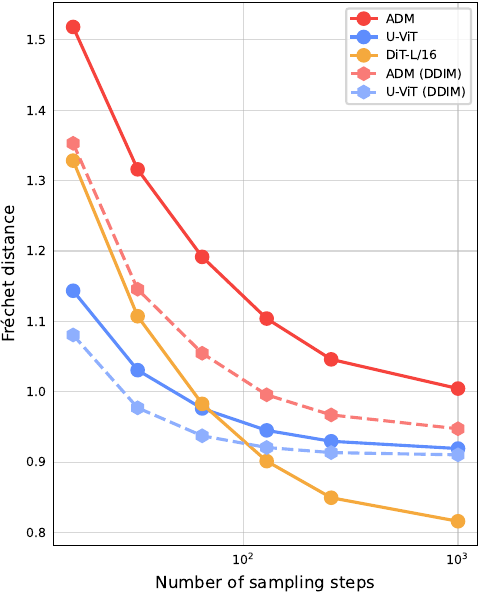}
    \caption{\textbf{Scaling-up sampling compute reduces FD score.} We compute the FD for using [16, 32, 64, 128, 256, 1000] sampling steps using both DDPM and DDIM sampling. We do not plot the DiT with DDIM sampling, as the FD score is overall too high and outside the plot to display properly.}
    \label{fig:fd_score}
\end{figure}

\begin{figure}[h]
    \centering
    \includegraphics{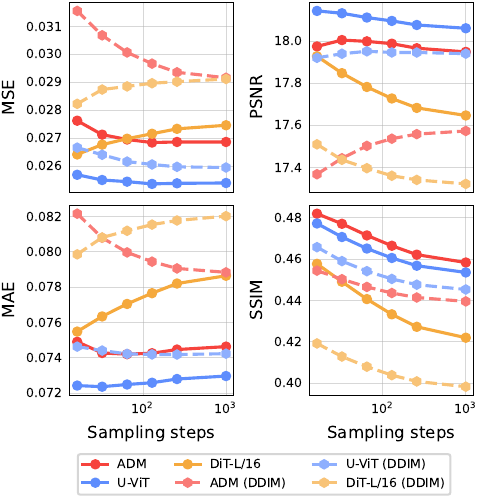}
    \caption{\textbf{Impact of increasing compute on intensity metrics.} For sampling steps [16, 32, 64, 128, 256, 1000] we compute the intensity based metrics MSE, MAE, PSNR and SSIM.}
    \label{fig:all_scores}
\end{figure}

We further compare the impact of different number of sampling steps in figures \ref{fig:fd_score} and \ref{fig:all_scores}. The U-ViT consistently outperforms the ADM in terms of FD, even for lower number of sampling steps (see figure \ref{fig:fd_score}). Moreover, the DiT-L/16 outperforms the U-ViT after 64 sampling steps. For all models, using 1000 sampling steps achieves the lowest FD score.

Using DDIM, all models improve in terms of FD score, except the DiT-L/16. When using 128 sampling steps we roughly reach the same FD with DDIM sampling as when using 1000 with DDPM sampling on the ADM and U-ViT. Towards 1000 sampling steps, both models seem to plateau, the U-ViT earlier than the ADM. For the ADM using DDIM sampling is consistently better in terms of FD score, even for 1000 sampling steps.

For the DiT we notice patch artefacts in all settings. They seem to be amplified / blurred when DDIM sampling is used. Using DDIM sampling for DiT-L/16 produces worse results. A simple fix for the patch artefacts we find, is to use a wavelet in a similar manner as for the transformer of the U-ViT. The model then operates in the latent space of the wavelet coefficients.

The use of more sampling steps consistently lowers the FD score, while for the intensity and image quality metrics, the effect varies depending on the metric and architecture. When sampling the ADM and U-ViT longer, the MSE improves or stagnates (see figure \ref{fig:all_scores}). For the DiT-L/16 the MSE score worsens the longer is sampled. Overall, of the diffusion models, the U-ViT has the best results for MSE, PSNR and MAE. The ADM has the highest score regarding SSIM. For MAE, PSNR and SSIM, sampling longer consistently worsens the performance. For 64 steps the ADM seems to reach the best performance on MAE and for 32 steps on PSNR.

\begin{table}[h]
\centering
\renewcommand{\arraystretch}{1.3}
\setlength\tabcolsep{12pt}
\caption{\textbf{Model compute.} For each model variant we report the number of parameters, throughput and estimate the giga floating point operations (GFLOPs) for a single forward pass (batch size of one) using JAX. Throughput is measured as iterations per second for bfloat16 where a single iteration is a batch of size 16 on a single NVIDIA A40 GPU.}
\begin{tabular}{@{}llll@{}}
\toprule
Model & \# Parameters & Throughput & GFLOPs \\ \midrule
U-Net & 15M & 14.14 it/s & 3 \\
ADM & 35M & 5.65 it/s & 22\\
U-ViT & 125M & 5.62 it/s & 120 \\
DiT-L/16 & 558M & 6.49 it/s & 46 \\
\bottomrule
\end{tabular}
\label{tab:model_compute}
\end{table}

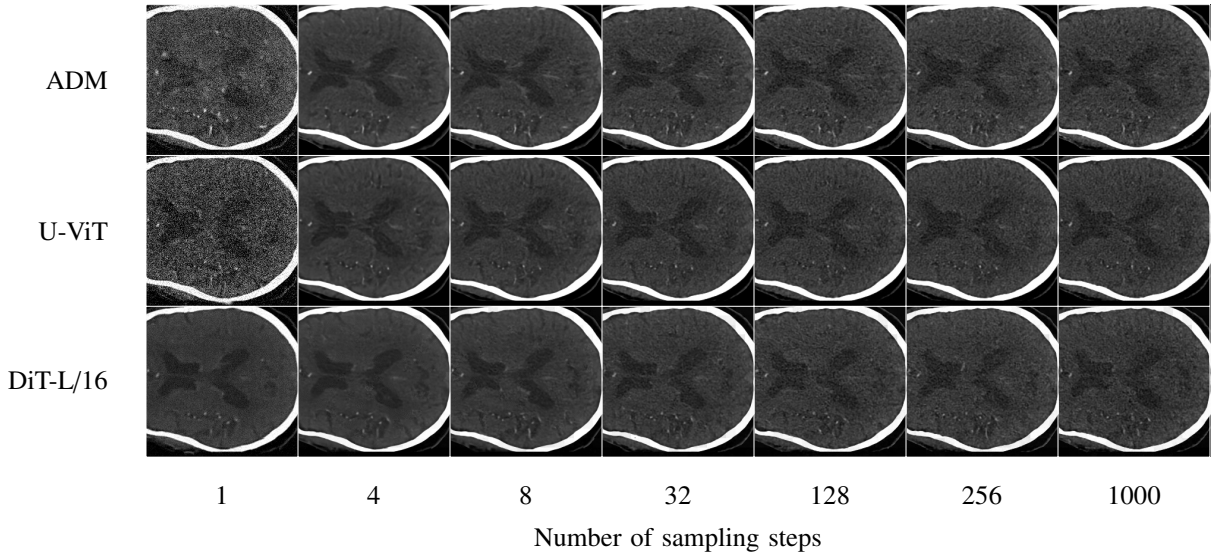
\begin{figure*}[h]
    \centering
    \begin{tikzpicture}
    \draw (0, 0) node[anchor=south west,inner sep=0] {\includegraphics[height=6cm, width=14cm]{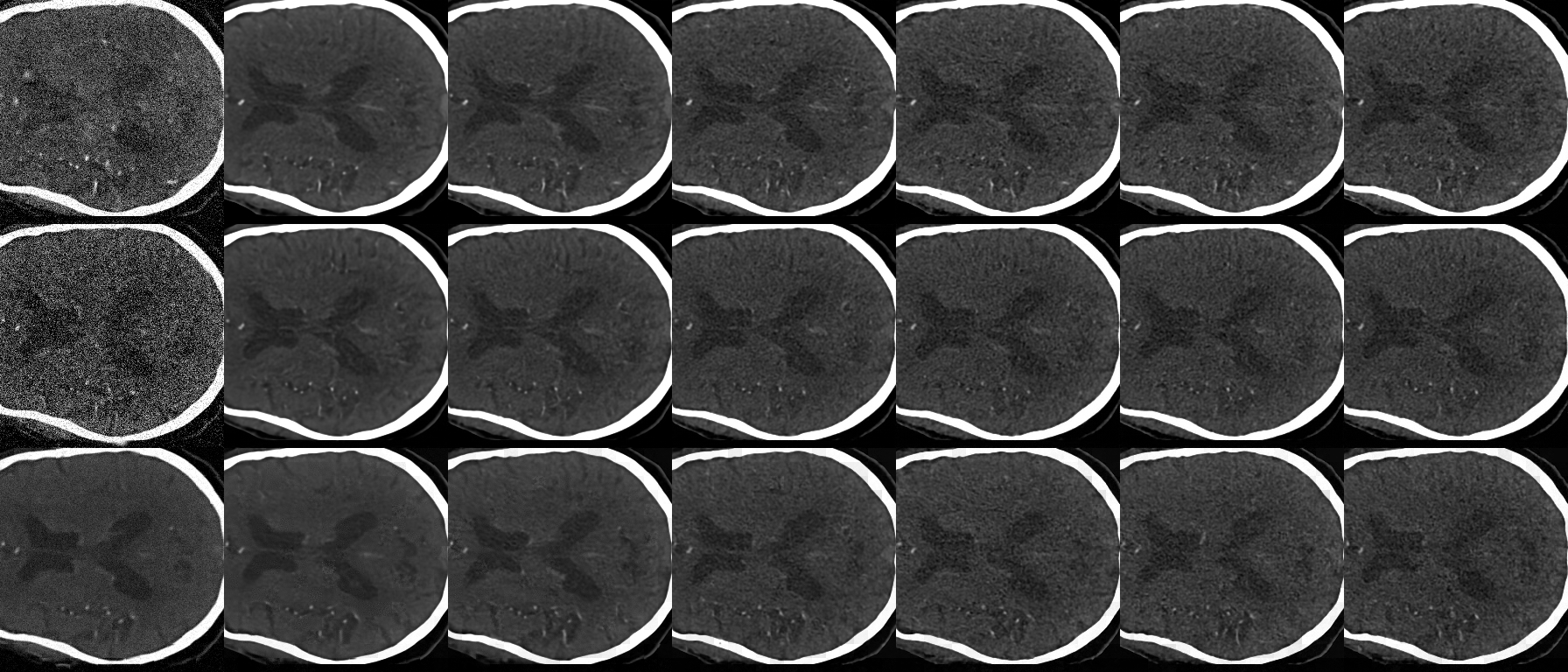}};
    \draw [help lines, step=2.0cm, color=white] (0,0) grid (14,6);
    \draw (1,-.5) node[align=center, text width=2cm] {1};
    \draw (3,-.5) node[align=center, text width=2cm] {4};
    \draw (5,-.5) node[align=center, text width=2cm] {8};
    \draw (7,-.5) node[align=center, text width=2cm] {32};
    \draw (9,-.5) node[align=center, text width=2cm] {128};
    \draw (11,-.5) node[align=center, text width=2cm] {256};
    \draw (13,-.5) node[align=center, text width=2cm] {1000};
    \draw (-1.5,5) node[align=right, text width=2cm] {ADM};
    \draw (-1.5,3) node[align=right, text width=2cm] {U-ViT};
    \draw (-1.5,1) node[align=right, text width=2cm] {DiT-L/16};
    \draw (7,-1.1) node[align=center, text width=8cm] {Number of sampling steps};
    \end{tikzpicture}
    \caption{\textbf{Comparison of number of sampling steps.} For each of the models: ADM, U-ViT and DiT, we plot different numbers of sampling steps $s \in [1,4,8,32,128,256,1000]$. We apply the default DDPM sampling.}
    \label{fig:sampling_steps_images}
\end{figure*}

\subsection{Volume synthesis}

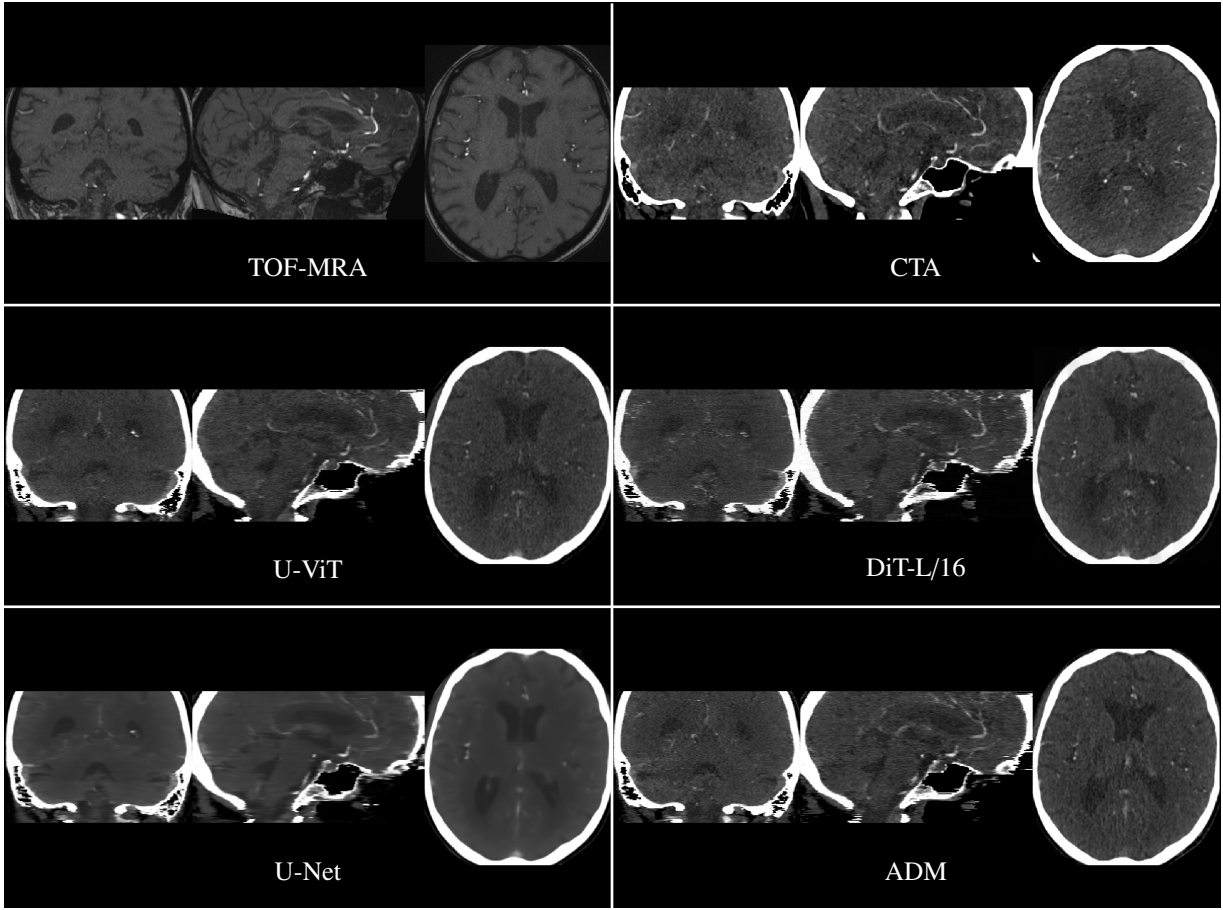
\begin{figure*}[h]
    \centering
    \begin{tikzpicture}
    \draw (0, 0) node[anchor=south west,inner sep=0] {\includegraphics[width=8cm, height=4cm]{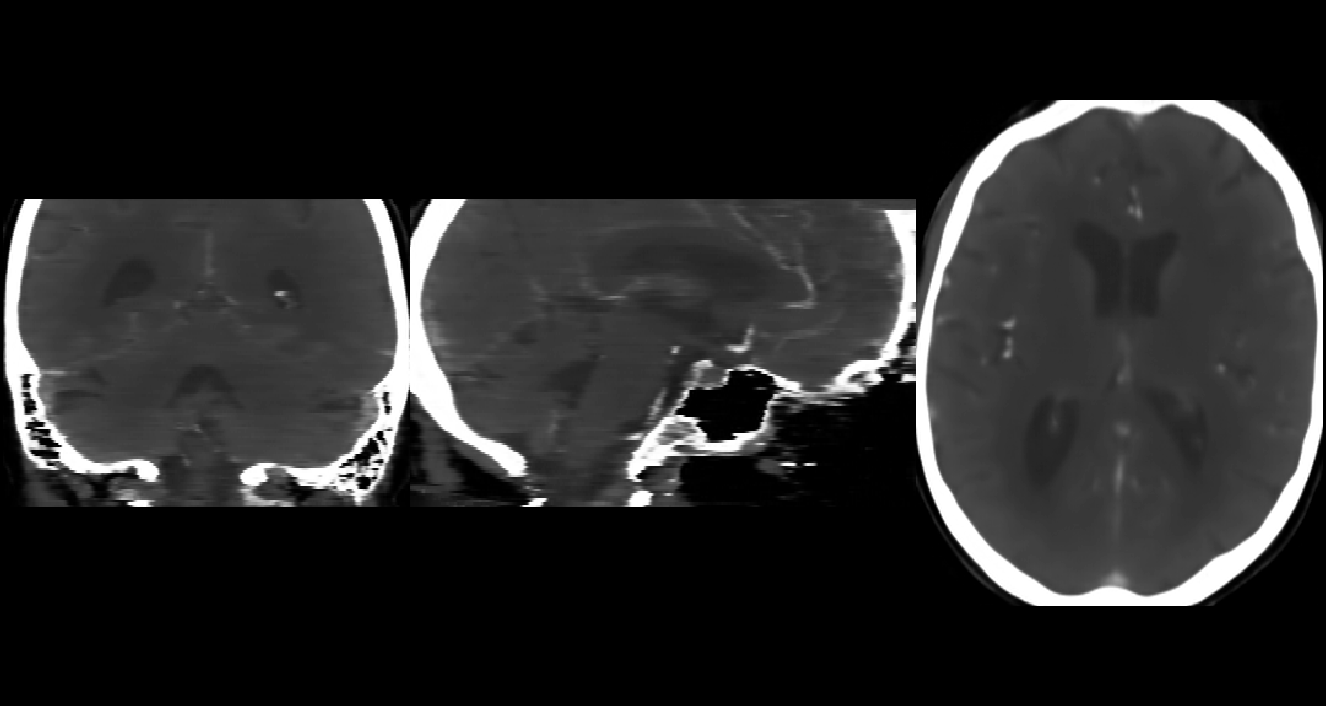}};
    \draw (4,.5) node[align=center, text width=2cm, color=white] {U-Net};
    
    \draw (8, 0) node[anchor=south west,inner sep=0] {\includegraphics[width=8cm, height=4cm]{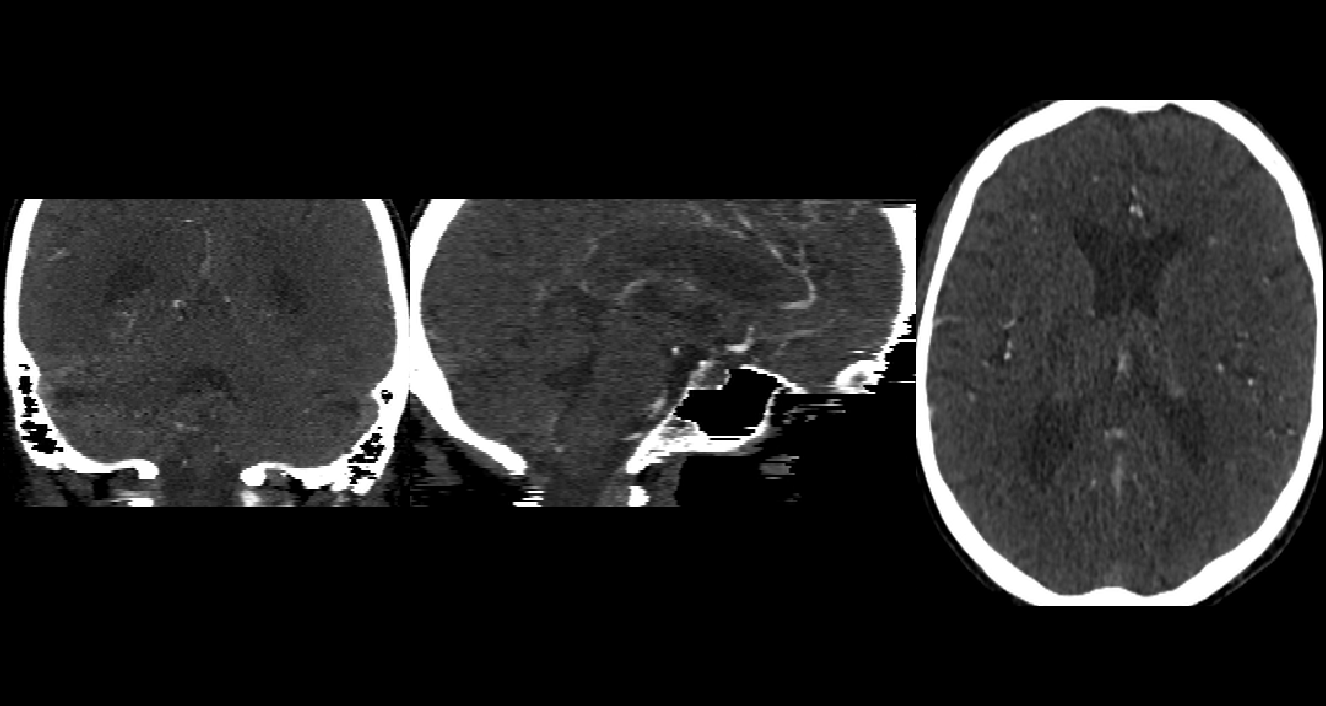}};
    \draw (12,.5) node[align=center, text width=2cm, color=white] {ADM};
    
    \draw (0, 4) node[anchor=south west,inner sep=0] {\includegraphics[width=8cm, height=4cm]{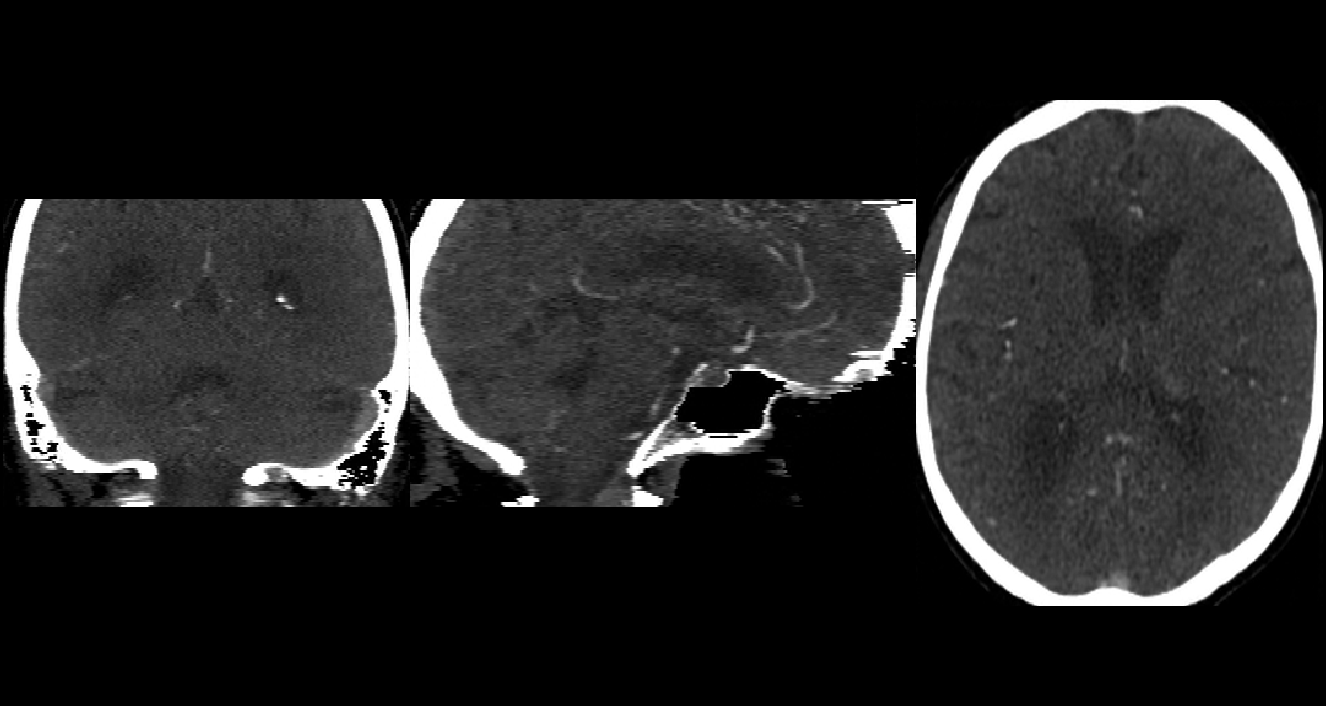}};
    \draw (4,4.5) node[align=center, text width=2cm, color=white] {U-ViT};
    
    \draw (8, 4) node[anchor=south west,inner sep=0] {\includegraphics[width=8cm, height=4cm]{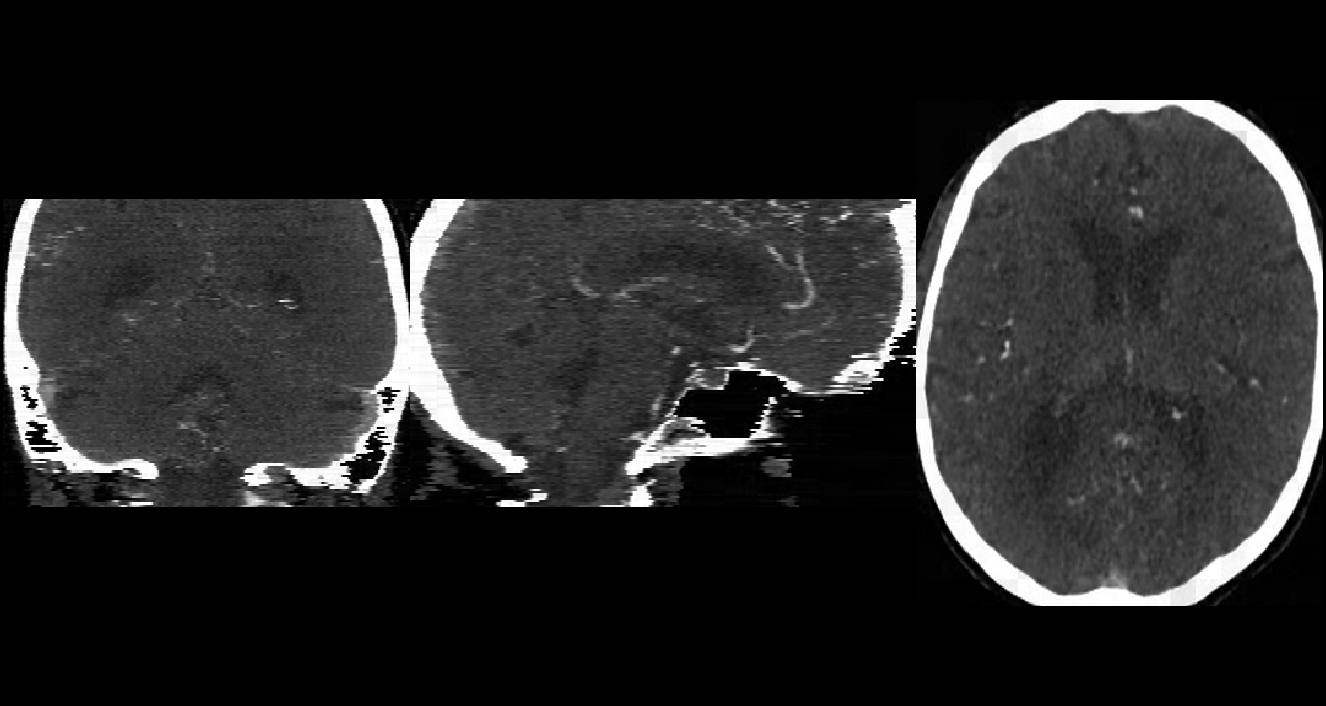}};
    \draw (12,4.5) node[align=center, text width=2cm, color=white] {DiT-L/16};

    \draw (0, 8) node[anchor=south west,inner sep=0] {\includegraphics[width=8cm, height=4cm]{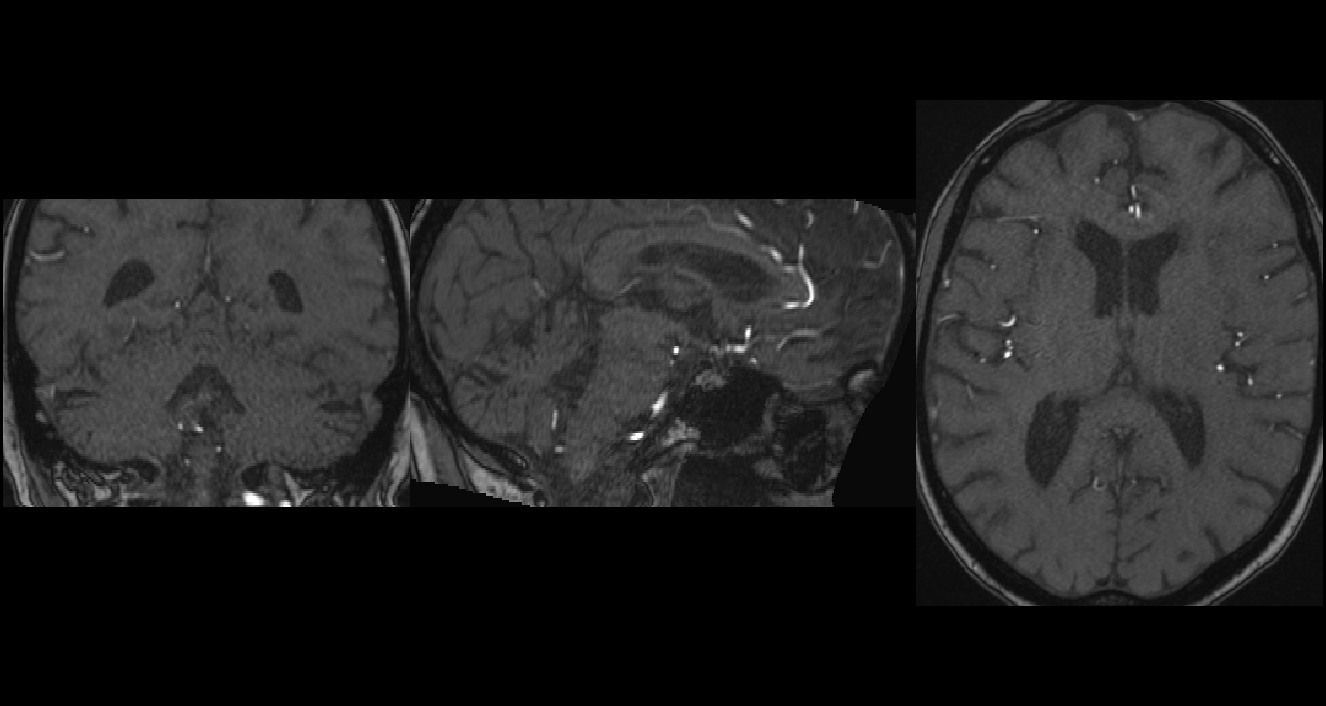}};
    \draw (4,8.5) node[align=center, text width=2cm, color=white] {TOF-MRA};
    
    \draw (8, 8) node[anchor=south west,inner sep=0] {\includegraphics[width=8cm, height=4cm]{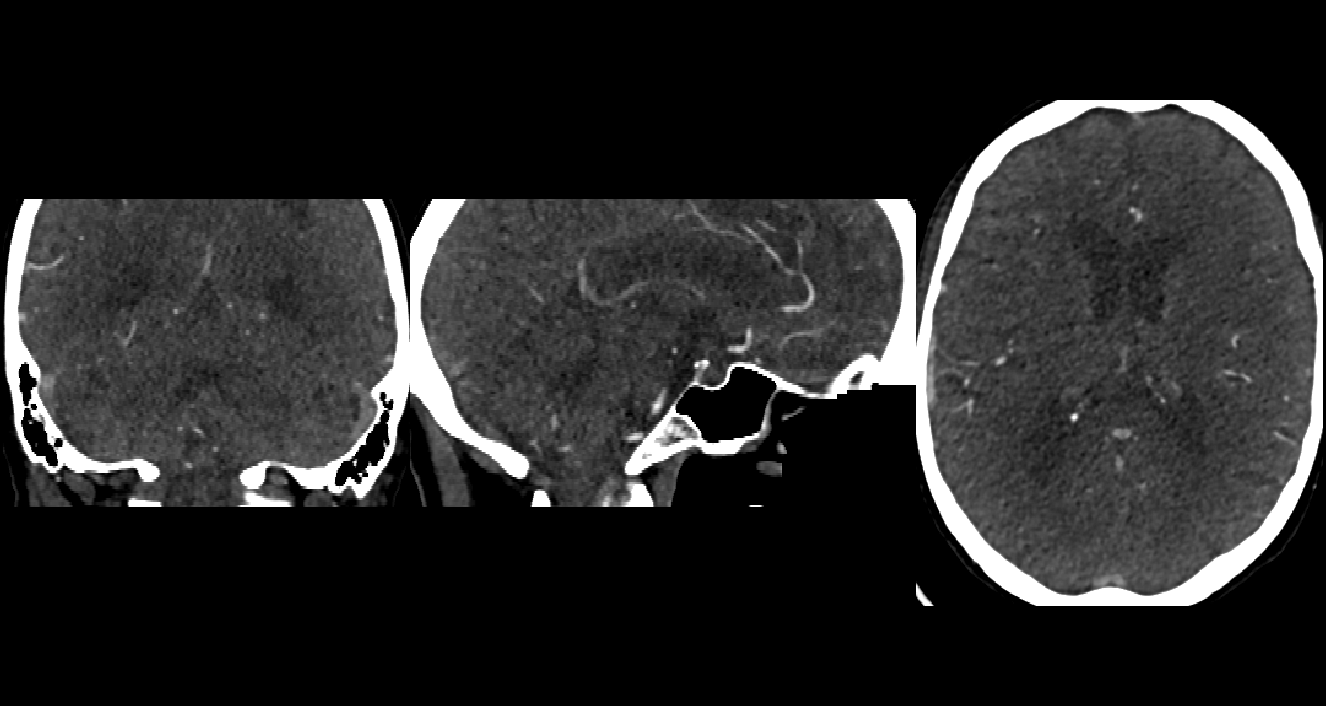}};
    \draw (12,8.5) node[align=center, text width=2cm, color=white] {CTA};

    \draw[white, line width=.1em] (0,4) -- (16,4);
    \draw[white, line width=.1em] (0,8) -- (16,8);
    \draw[white, line width=.1em] (8,0) -- (8,12);
    \end{tikzpicture}
    
    \caption{\textbf{Full brain plots.} A sample from our TopCoW data test set is reconstructed in 3D using the proposed resampling method. We apply 128 DDPM sampling steps. Top left: Source TOF-MRA image, top right: Target CTA image, middle left: U-ViT, middle right: DiT-L/16, bottom left: U-Net, bottom right: ADM. Due to the defacing of the ground truth CTA images, artefacts appear in the synthetically generated images.}
    \label{fig:full_brains}
\end{figure*}

\begin{table}[h]
\centering
\renewcommand{\arraystretch}{1.4}
\setlength\tabcolsep{3pt}
\caption{\textbf{Volume synthesis.} Results of generating the full brain volumes on the TopCoW dataset. Metrics are computed on $[-50,350]$ range. SSIM is computed using a 3D Gaussian blur kernel with window size of 11. We use DDPM sampling with 128 sampling steps. FD score is calculated slice-wise on the transverse plane. Underlined are the second best results.}
\scalebox{0.85}{
\begin{tabular}{@{}lllllll@{}}
\toprule
Model & Sampler & MSE ($\downarrow$) & MAE ($\downarrow$) & PSNR ($\uparrow$) & SSIM ($\uparrow$) & FD ($\downarrow$)  \\ \midrule
U-Net & N/A & \textbf{9024.022} & \textbf{43.309} & \textbf{12.525} & \textbf{0.456} & 12.040 \\ \cmidrule{1-7}
ADM & DDPM & 10158.872 & 45.334 & 11.998 & 0.397 & 5.953 \\
U-ViT & DDPM & 10133.412 & \underline{45.303} & 12.007 & \underline{0.405} & \textbf{5.753} \\
DiT-L/16 & DDPM & \underline{9760.138} & 45.982 & \underline{12.184} & 0.388 & 5.874 \\ \cmidrule{1-7}
ADM & DDIM & 10350.404 & 46.332 & 11.918 & 0.383 & 5.992 \\
U-ViT & DDIM & 10164.413 & 45.719 & 11.995 & 0.399 & \underline{5.827} \\
DiT-L/16 & DDIM & 9879.507 & 46.604 & 12.131 & 0.373 & 6.247 \\
\bottomrule
\end{tabular}}
\label{tab:volume_eval_internal}
\end{table}

For the volume synthesis experiments we use DDPM sampling with 128 sampling steps. This is fast enough while providing good sample quality. In figure \ref{fig:full_brains} we show one sample of the test set and the results of the generation process using all of our models. Since the method is two-dimensional only, one can notice slice artefacts, where the model is uncertain about where the skull begins and the brain ends. Moreover, due to the anonymization i.e.\ the defacing of the images, the model is uncertain about removing the face.

In table \ref{tab:volume_eval_internal} we show the results of the model evaluation of the full volume. We perform reconstruction using our proposed slice-wise resampling method. The evaluation is performed in CTA space, both images are rescaled and clipped to $[-50,350]$ range. We calculate the FD score slice-wise on the transverse plane and average the results per patient. The U-Net achieves the lowest scores on MSE ($9024.022$), MAE ($43.309$), PSNR ($12.525$) and SSIM ($0.456$), while the U-ViT achieves the lowest FD score ($5.753$). Furthermore, the U-ViT has the second best scores for SSIM ($0.405$) and MAE ($45.303$). For MSE and PSNR the DiT achieves the second best scores of $9760.138$ and $12.184$ respectively.

\subsection{External validation}
\begin{table}[h]
\centering
\renewcommand{\arraystretch}{1.4}
\setlength\tabcolsep{3pt}
\caption{\textbf{External data volume synthesis.} Results of generating the full brain volumes on external test set. Metrics are computed on $[-50,350]$ range. SSIM is computed using a 3D Gaussian blur kernel with window size of 11. FD score is calculated slice-wise on the transverse plane. We use DDPM sampling with 128 sampling steps. Underlined are the second best results.}
\scalebox{0.85}{
\begin{tabular}{@{}lllllll@{}}
\toprule
Model & Sampler & MSE ($\downarrow$) & MAE ($\downarrow$) & PSNR ($\uparrow$) & SSIM ($\uparrow$) & FD ($\downarrow$)  \\ \midrule
U-Net & N/A & 11805.073 & 61.204 & 11.370 & 0.306 & 14.581 \\ \cmidrule{1-7}
ADM & DDPM & \textbf{10479.845} & \textbf{50.083} & \textbf{11.950} & \underline{0.358} & 9.089 \\
U-ViT & DDPM & 11733.429 & \underline{52.910} & 11.440 & \textbf{0.383} & \textbf{8.482} \\
DiT-L/16 & DDPM & 13721.648 & 60.835 & 10.818 & 0.333 & 9.778 \\ \cmidrule{1-7}
ADM & DDIM & \underline{11556.339} & 54.979 & \underline{11.494} & 0.304 & 9.359 \\
U-ViT & DDIM & 12120.984 & 55.286 & 11.287 & 0.351 & \underline{8.774} \\
DiT-L/16 & DDIM & 14205.849 & 62.756 & 10.660 & 0.312 & 9.671 \\
\bottomrule
\end{tabular}
}
\label{tab:volume_eval_external}
\end{table}
In table \ref{tab:volume_eval_external} we show the metric differences of the models on the external test set using the resampling method. Overall all models have slightly worse performance. The external imaging data was difficult to register, thus many images are missing parts in the TOF-MRA imaging compared to the CTA data.

Compared to the internal validation in table \ref{tab:volume_eval_internal}, where the U-Net excels on the metrics, on the external set we observe different results. Here the ADM performs best for MSE ($10479.845$), MAE ($50.083$) and PSNR ($11.95$). The U-ViT has the highest SSIM of $0.383$ and the lowest FD of $8.482$. The second-best model is also always a diffusion model for all metrics. For MSE, PSNR and SSIM the second-best model is again the ADM. The second-best MAE of $52.910$ is achieved by the U-ViT.

\section{Discussion}

In this paper we compare different diffusion models that operate on a slice-wise level to perform image-to-image translation. Our results show, that on the TopCoW data, the U-Net performs best based on the MSE, MAE, PSNR and SSIM scores. For 3D reconstruction on TopCoW we see similar results. In contrast, for the task of 3D reconstruction on the external data, the diffusion models perform better than the U-Net. Overall, images generated by the U-Net appear blurry, while the diffusion models include more fine-grained details. The FD score, which has been found to be indicative of human perception, agrees with this, since it is lower for all diffusion models than that of the U-Net. On all three tasks, slice-wise prediction, internal and external reconstruction, the diffusion models have the lowest FD score. Therefore, we argue that the diffusion models generate more realistic results than a standard U-Net.

Our observations show that standard metrics for image-to-image translation do not align well with the human qualitative assessment of synthetic images. While MSE, MAE, PSNR and SSIM are standard metrics for image-to-image translation tasks, we would also like to highlight that MSE and PSNR do not capture blurring \citep{ndajah20ssim}. Moreover, PSNR and SSIM are highly sensitive to rotations, spatial shifts and scaling \citep{wang09mse}, as well as Gaussian noise \citep{kotevski10psnr}. CTA imaging as a modality is highly noisy is contrast to the TOF-MRA imaging. Therefore, we argue that the metrics fail to be a good method for predicting the perceived quality of the generated images. MSE, MAE and PSNR are reporting absolute errors and are thus prone to slight changes in the image density. We hypothesize that the perceived change in structural information is somewhat captured by SSIM, but due to spatial shifts, i.e.\ the uncertainty of the diffusion model and the residual noise left in the image, the score must favour the standard U-Net-based result. In figure \ref{fig:appendix_metrics_schedule_example}, we show that by applying different noise schedules, the diffusion models can outperform the results of the U-Net metric-wise, while creating unrealistic results.

Since the best performing model differs on the TopCoW dataset and the external test set, we recommend trying the different diffusion-models and visually inspecting which produce the best results for the application on new datasets. Because the ADM performs best on MSE, MAE and PSNR the external test set, we suggest trying it as a good starting point. However, the U-ViT achieves the best SSIM, which might suggest better overall image quality. If speed is an issue, we recommend the DiT, as it is the fastest model. If one wants to reduce the number of sampling steps, we recommend using DDIM sampling on the ADM and U-ViT models, as DDIM sampling consistently creates lower FD scores for these models than DDPM sampling, at a negligible cost of worsening the other metrics. When running DDPM sampling long enough it should be preferred over DDIM as it creates more accurate results.

When applying the models to images of new sites, we found that the images should be tightly cropped, containing the entire skull, to produce the best results. Further, for images of high resolution (i.e.\ $512 \times 512$), using the ADM and U-ViT directly without resampling produces better results. The DiT fails to work on high resolutions, due to the bad extrapolation ability of the sinusoidal positional embedding.

\section{Limitations}

Our work has several limitations. A first possible downside of the application of diffusion models is the long sampling time. Sampling a single slice takes a few seconds up to minutes, depending on the slice resolution. Performing this for all slices of a brain image volume can be time-consuming. This limits the clinical application, especially if no cluster-grade GPUs are available. In this paper we limit ourselves to sampling using DDPM or DDIM. While DDIM sampling creates better results for lower number of function evaluations, we do not report results on more recent diffusion solvers. Thus, to obtain optimal results, long sampling times are required.

Second, we did not explore three-dimensional generation as this requires exponentially more samples and diffusion models are data hungry. CT images are large in size and would need to be significantly compressed (e.g.\ using an Autoencoder) for our model to work, given the GPU memory limitations. We did not train a latent diffusion model because it is likely to overfit on the limited data (on both the Autoencoder and the diffusion model). Our two-dimensional models also do not take neighbouring slices into account. Exploring such a 2.5D approach we leave for future work.

Third, we did not employ more extensive hyperparameter tuning and regularization. Using specialized augmentations and regularization in combination with learning rate scheduling could likely further improve the robustness and generalization performance of the models. Using exponential moving averages of the weights during training has become a standard practice for diffusion models, however, for our models, we did not see improvements. Moreover, since the models are very time-expensive to train and evaluate we did not perform multiple reruns per model on different model initialization seeds to provide standard deviations on the provided metrics.

\section{Conclusions and Future Work}

We showed a promising way of performing modality conversion in vessel neuroimaging. The proposed method can be used to generate synthetic CTA data from available TOF-MRA datasets. Synthetic CTA images can be beneficial in several ways. Existing TOF-MRA imaging datasets can be augmented by synthesizing the corresponding CTAs, bypassing resource-intensive data transfer, acquisition or pre-processing steps such as registration. The availability of CTA images in addition to TOF-MRAs can enable models to incorporate anatomical information jointly from both modalities. While not assessed in our work, synthetic data was shown to improve fairness and generalization of classification models in medical models \citep{Ktena2024}. Synthetic CTA data can be used in downstream AI applications such as aneurysm segmentation, occlusion detection and automated collateral score assessment. This has the potential to improve the accuracy and generalization of AI models in medical imaging. The question, whether synthetic CTAs can retain the diagnostic superiority of CTAs in certain conditions remains open. Future work should assess the diagnostic accuracy of radiologists in clinical tasks such as aneurysm detection or large vessel occlusion (LVO) detection, when synthetic CTAs are provided as additional imaging information. The potential benefits and use cases of synthetic CTA images in clinical practice could be explored further.

Future work could also explore technical improvements. Different noise schedules should be explored, as we found this has a greater impact on generalization than initially thought. Shifting the cosine noise schedule according to a reference resolution as proposed in \cite{hoogeboom_simple_2023} can be explored as well as sigmoid noise schedules \cite{jabri23rin,chen23noiseschedules}. The sampling efficiency could also be improved, by implementing other diffusion solvers, using progressive distillation \citep{salimans_progressive_2022} or using consistency models \citep{song23consistency,heek24}, which enable single-shot sampling. Moreover, architectural changes, such as proposed by \cite{karras_analyzing_2024,crowson24hdit} can further improve the quality of generated images. Additionally, providing more slices as conditioning could improve the inter-slice robustness of the model for 3D sampling. Finally, better evaluation metrics should be found which incorporate both structural similarity and perceived texture.

\section*{Acknowledgements}

The authors acknowledge the financial support by the Federal Ministry of Education and Research of Germany in the grant program \enquote{Forschungsnetzwerk Anonymisierung für eine sichere Datennutzung} (Project number 16KISA042K).
Computation has been performed on the HPC for Research cluster of the Berlin Institute of Health.

\section*{CRediT authorship contribution statement}

\textbf{Alexander Koch:} Conceptualization, Data curation, Formal analysis, Investigation, Methodology, Software, Validation, Visualization, Writing – original draft, Writing – review and editing. \textbf{Orhun Utku Aydin:} Conceptualization, Data curation, Writing – original draft, Writing - Review \& Editing.
\textbf{Adam Hilbert:} Writing - Review \& Editing.
\textbf{Jana Rieger:} Writing - Review \& Editing.
\textbf{Satoru Tanioka:} Data curation, Writing - Review \& Editing.
\textbf{Fujimaro Ishida}: Data curation.
\textbf{Dietmar Frey}: Resources, Funding acquisition, Writing - Review \& Editing.

\section*{Data availability}

The TopCoW challenge data is publicly available through \url{https://topcow23.grand-challenge.org/data/}. Our private external test set can not be shared at this time.

\bibliographystyle{elsarticle-harv.bst}\biboptions{authoryear}
\bibliography{references}

\appendix

\onecolumn

\section{Changing the noise schedule post-hoc}

Since our diffusion models are trained on continuous-time diffusion, and they only require a valid log-space SNR value, we can simply swap out the noise schedule. Since the images are quite large in size $(256 \times 256)$, it is recommended to shift the schedule so that enough noise is added to the image \citep{hoogeboom_simple_2023,chen23noiseschedules}. The shifted cosine schedule is defined as

\begin{align}
    \log \mathrm{SNR}(t) &= -2 \log \tan (\pi t / 2) + 2 \log (d / 256)
\end{align}

where $256 \times 256$ is the image resolution and $d \times d$ is the reference image resolution. \citet{jabri23rin} propose a sigmoid-based noise schedule parameterized by $s,e,\tau$.

\begin{figure*}[h]
    \centering
    \includegraphics[width=18cm]{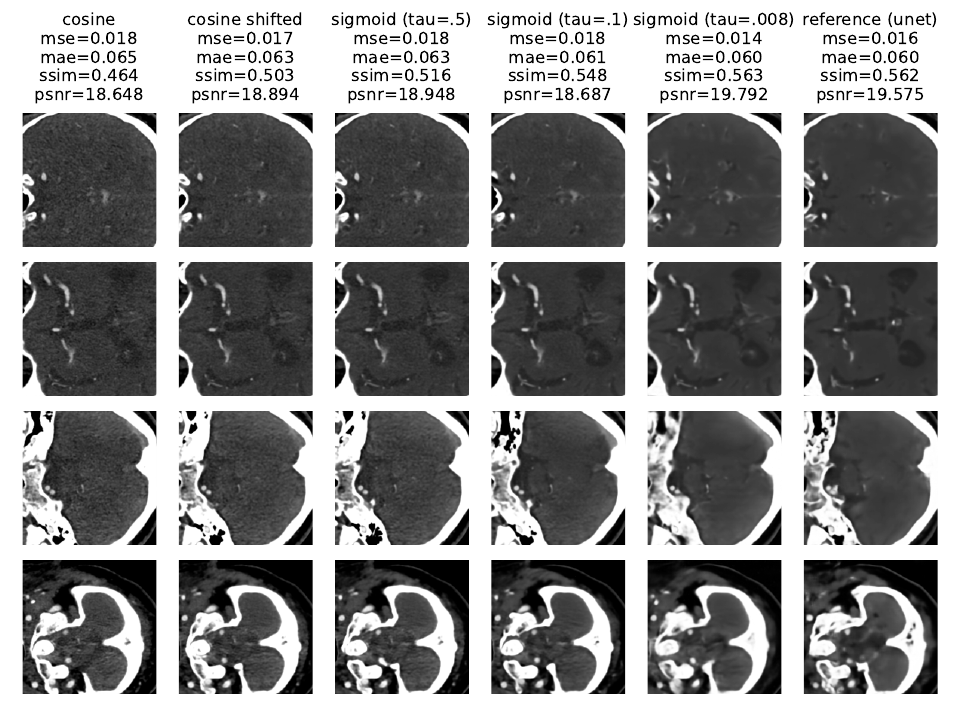}
    \caption{\textbf{Metrics for different noise schedules.} We show the ADM with different noise schedules. The leftmost column shows the normal cosine noise schedule. The next column shows a shifted cosine schedule, followed by three different sigmoid noise schedules with different scales. The last column shows the reference U-Net generated results. The metrics are averaged for the entire batch of four images. The images are drawn from the TopCoW test set. We sample using 128 sampling steps with DDPM sampling.}
    \label{fig:appendix_metrics_schedule_example}
\end{figure*}

In figure \ref{fig:appendix_metrics_schedule_example} we show the results of swapping out the noise schedule on the ADM on a single batch of four of the test set. We choose five different schedules: a cosine schedule, a shifted cosine schedule and three sigmoid schedules with different strengths. Each schedule progressively adds more noise, i.e.\ the noise level is high for more steps during sampling. For the shifted cosine schedule we use a reference resolution of 32. We find that when using these schedules, the metrics MSE, MAE, PSNR and SSIM improve. We can even outperform the U-Net slightly using one of the sigmoid schedules. However, the resulting images resemble more and more the U-Net's and appear less realistic.

\section{Additional Images}

\begin{figure*}[h]
    \centering

    \begin{tikzpicture}
    \draw (0, 0) node[anchor=south west,inner sep=0] {\includegraphics[width=8cm, height=4cm]{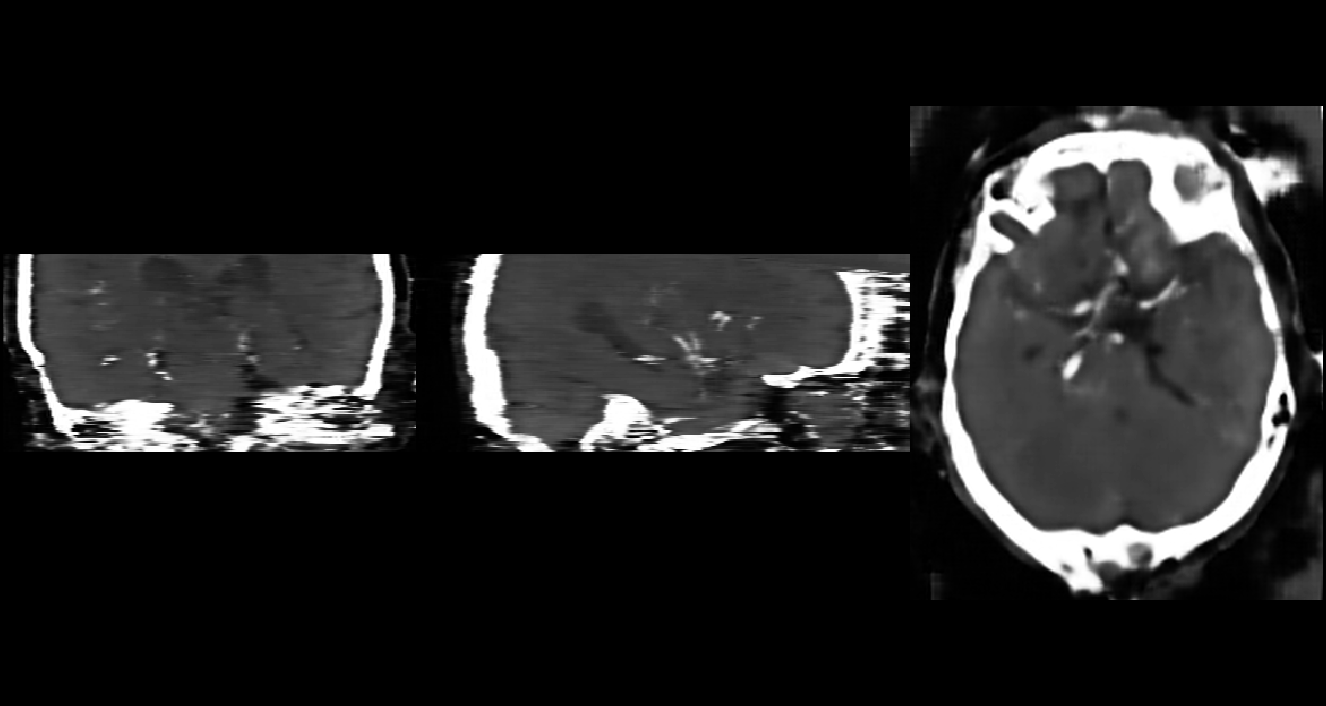}};
    \draw (4,.5) node[align=center, text width=2cm, color=white] {U-Net};
    
    \draw (8, 0) node[anchor=south west,inner sep=0] {\includegraphics[width=8cm, height=4cm]{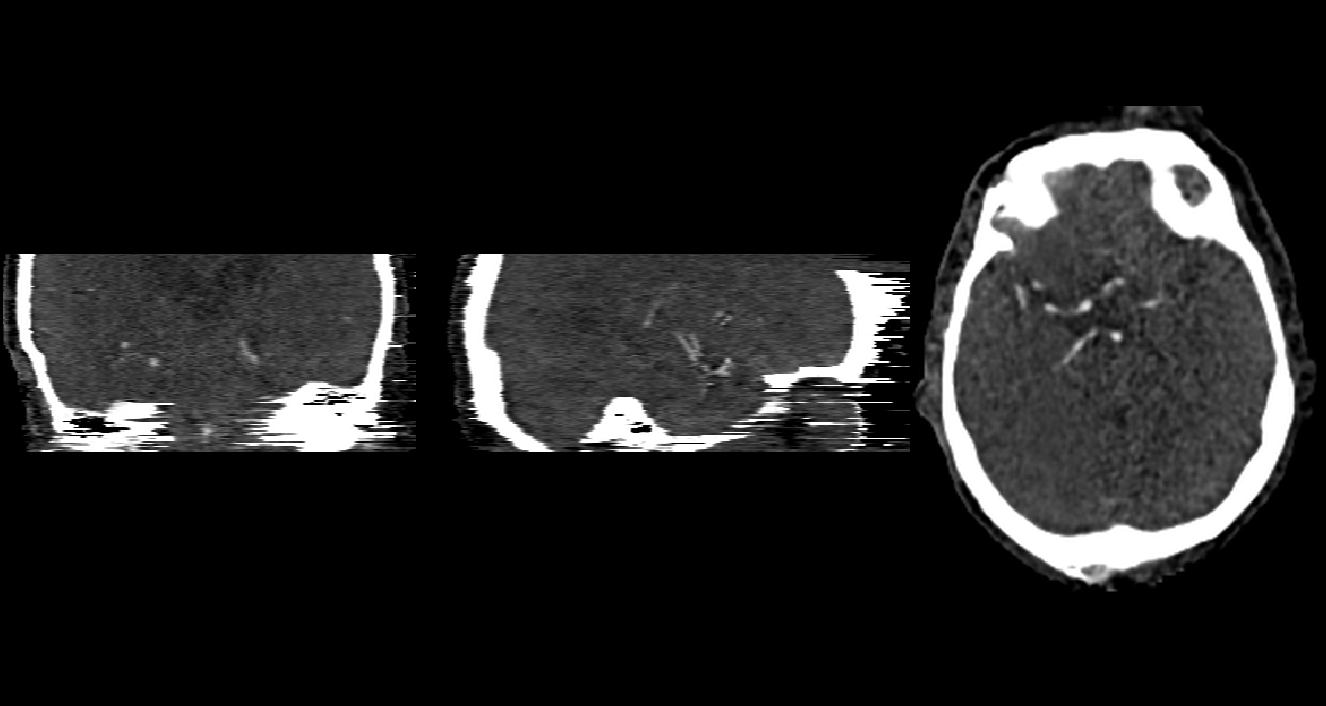}};
    \draw (12,.5) node[align=center, text width=2cm, color=white] {ADM};
    
    \draw (0, 4) node[anchor=south west,inner sep=0] {\includegraphics[width=8cm, height=4cm]{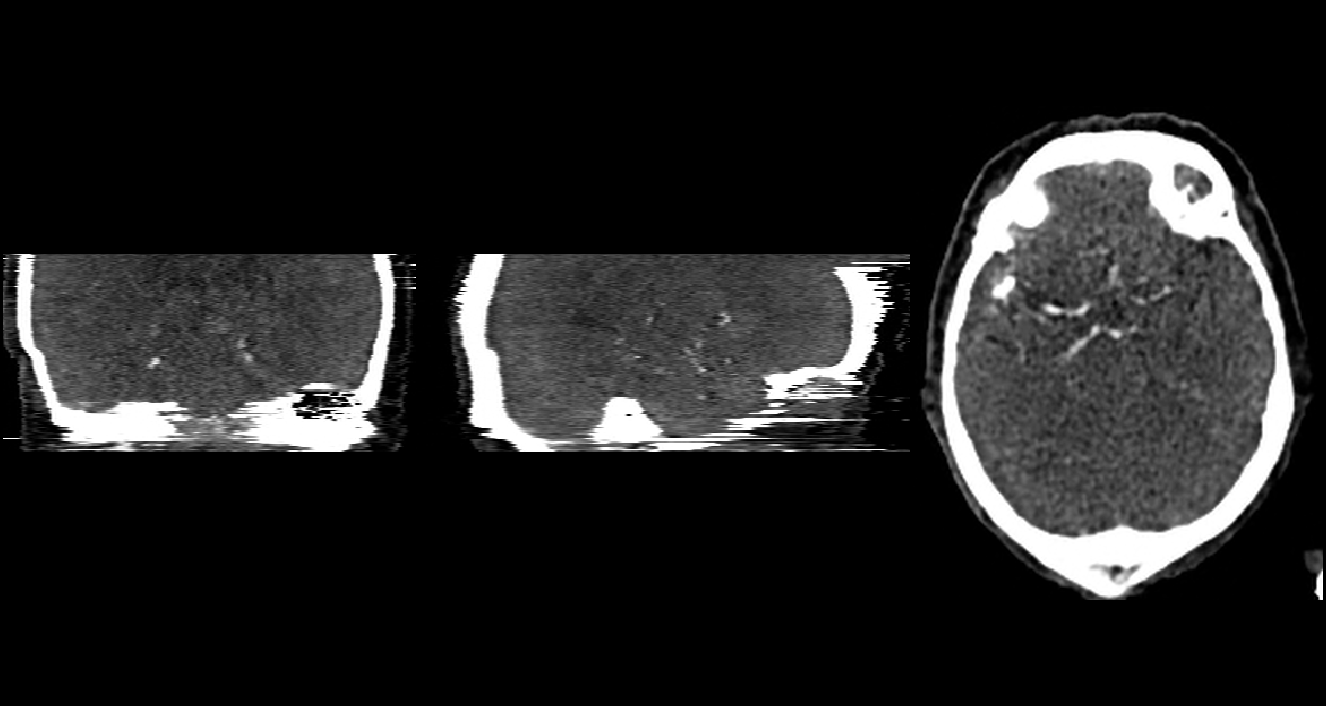}};
    \draw (4,4.5) node[align=center, text width=2cm, color=white] {U-ViT};
    
    \draw (8, 4) node[anchor=south west,inner sep=0] {\includegraphics[width=8cm, height=4cm]{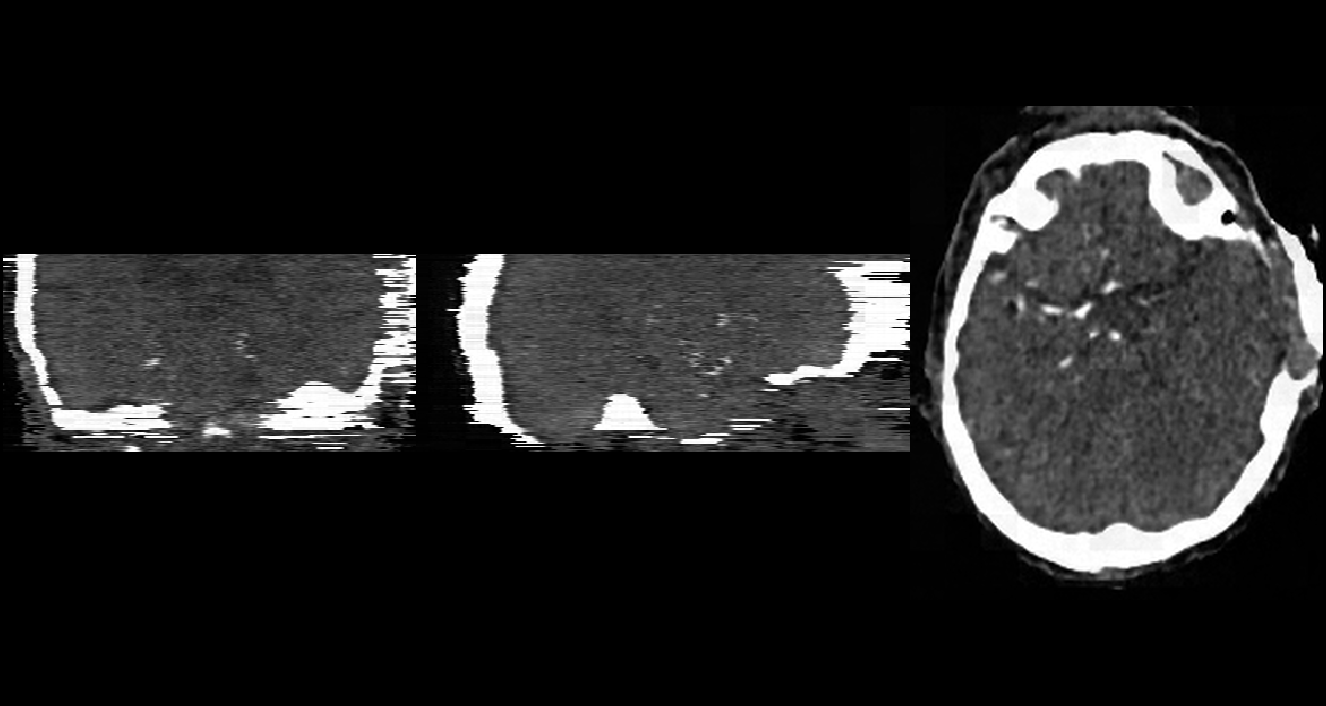}};
    \draw (12,4.5) node[align=center, text width=2cm, color=white] {DiT-L/16};

    \draw (0, 8) node[anchor=south west,inner sep=0] {\includegraphics[width=8cm, height=4cm]{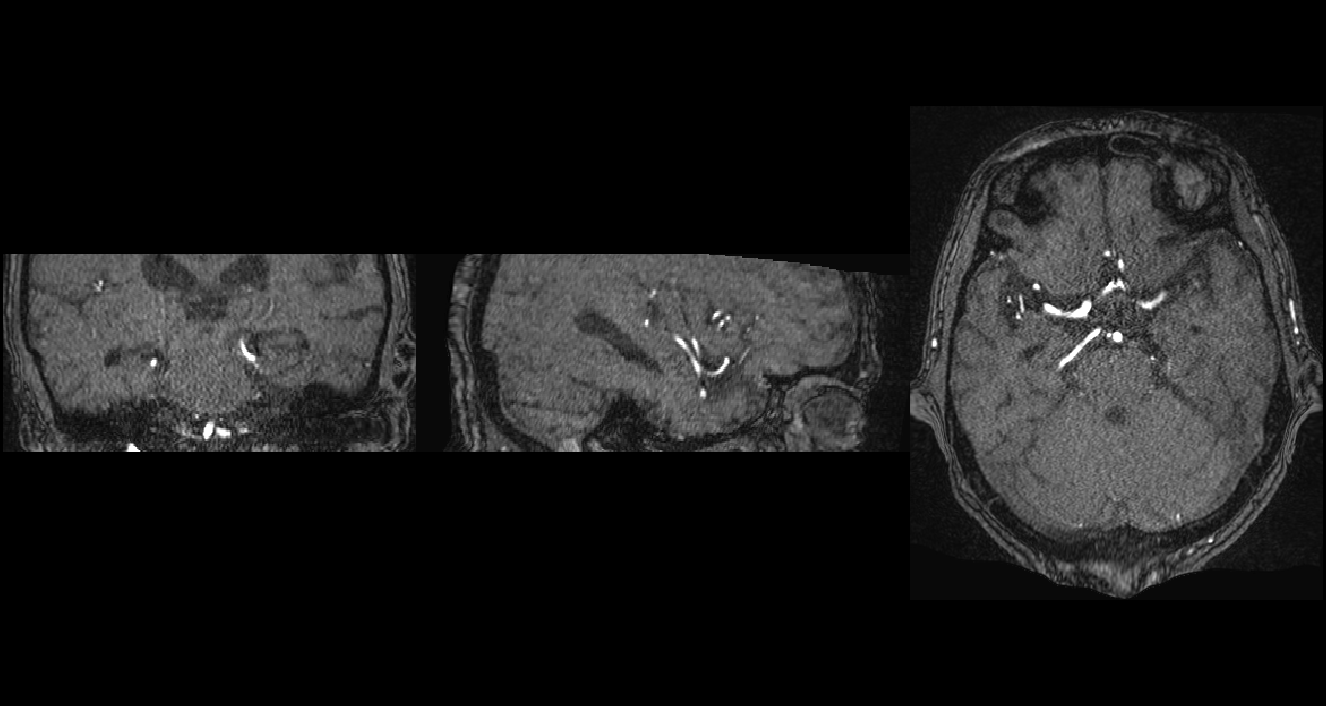}};
    \draw (4,8.5) node[align=center, text width=2cm, color=white] {TOF-MRA};
    
    \draw (8, 8) node[anchor=south west,inner sep=0] {\includegraphics[width=8cm, height=4cm]{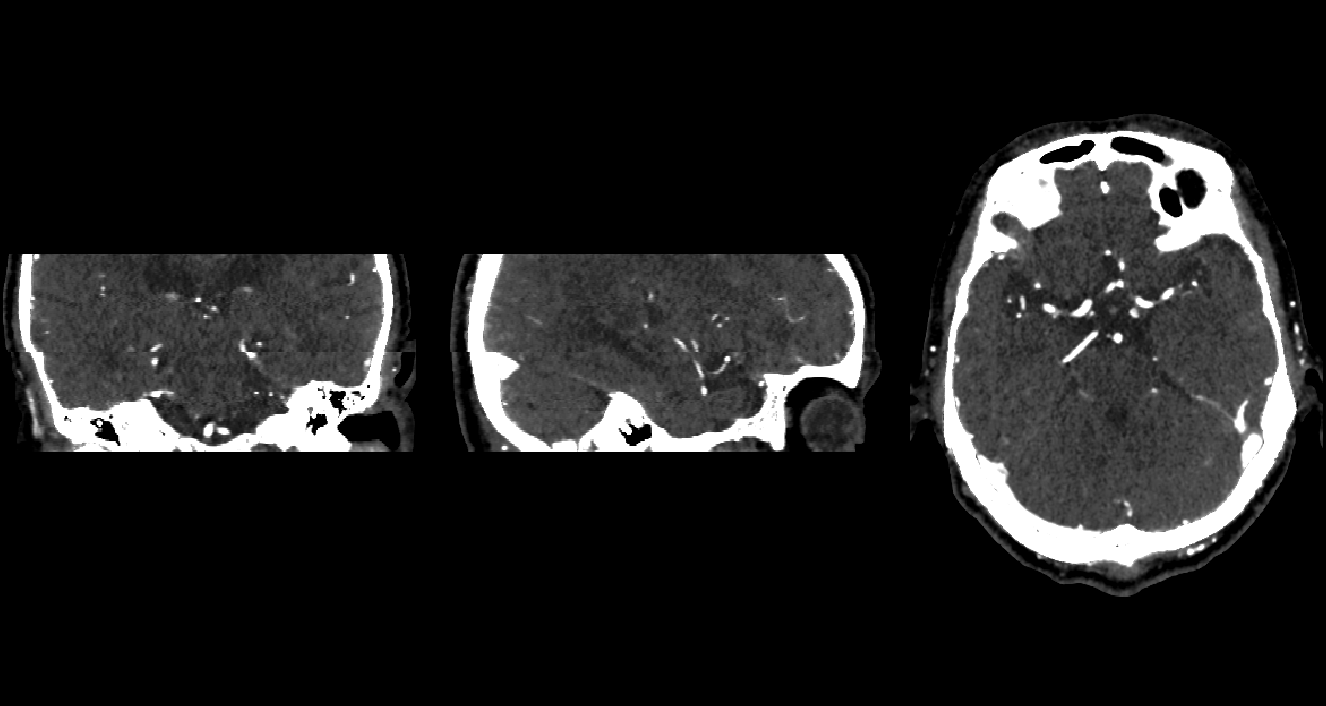}};
    \draw (12,8.5) node[align=center, text width=2cm, color=white] {CTA};

    \draw[white, line width=.1em] (0,4) -- (16,4);
    \draw[white, line width=.1em] (0,8) -- (16,8);
    \draw[white, line width=.1em] (8,0) -- (8,12);
    \end{tikzpicture}
    
    \caption{\textbf{External full brain plot.} A sample from our external test set, reconstructed in 3D using the proposed resampling method. We use 128 sampling steps with DDPM sampling.}
    \label{fig:appendix_external_full_brains}
\end{figure*}
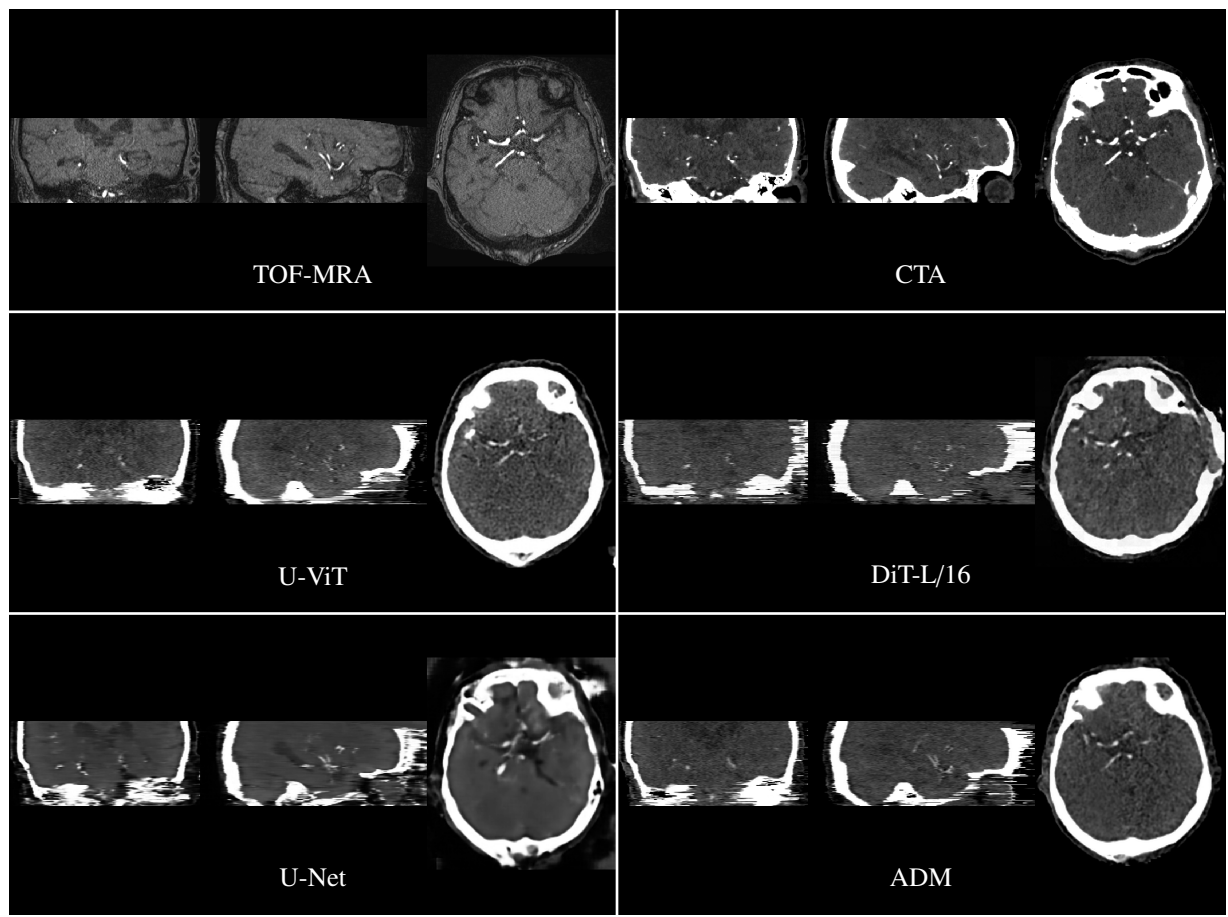

In this section we present an additional full brain sample from the external test set in figure \ref{fig:appendix_external_full_brains}.

\newpage

\section{Imaging parameters}
\begin{table}[h]
    \centering
    \renewcommand{\arraystretch}{2.5}
    \setlength\tabcolsep{4pt}
\scalebox{0.48}{
\begin{tabular}{@{}lllllllllllll@{}}
\toprule
Scanner Manufacturer & Scanner Model & Modality & Pixel Spacing (mm) & Slice Thickness (mm) & Patient ID & Acquisition Time & kVp & Exposure Time (ms) & X-ray Tube Current (mA) & Exposure (mAs) & Body Part Examined & Magnetic Field Strength (T)\\ \midrule
SIEMENS & nan & CT & [0.391,0.391] & 0.600 & ZA84e79311 & 120446.438 & 80.000 & 570.000 & 350.000 & 200.000 & HEAD & nan\\
SIEMENS & nan & MR & [0.562,0.562] & 0.640 & ZA84e79311 & 121802.152 & nan & nan & nan & nan & HEAD & nan\\
TOSHIBA & Aquilion ONE & CT & [0.468,0.468] & 0.500 & ZA05d74016 & 110935.050 & 80.000 & 1000.000 & 100.000 & 100.000 & HEAD & nan\\
Philips Medical Systems & Ingenia CX & MR & [0.396,0.396] & 1.350 & ZA05d74016 & 100338.290 & nan & nan & nan & nan & BRAIN & nan\\
SIEMENS & nan & CT & [0.391,0.391] & 0.600 & ZA41200276 & 123935.627 & 80.000 & 570.000 & 350.000 & 200.000 & HEAD & nan\\
SIEMENS & nan & MR & [0.562,0.562] & 0.640 & ZA41200276 & 125849.130 & nan & nan & nan & nan & HEAD & nan\\
SIEMENS & nan & CT & [0.400,0.400] & 0.500 & ZAe9853ae0 & 130420.080 & 100.000 & 500.000 & 561.000 & 510.000 & HEAD & nan\\
SIEMENS & nan & MR & [0.292,0.292] & 0.600 & ZAe9853ae0 & 110936.340 & nan & nan & nan & nan & HEAD & nan\\
GE MEDICAL SYSTEMS & nan & CT & [0.449,0.449] & 625.000 & ZAe7b228f4 & 223519.354 & 80.000 & 400.000 & 180.000 & 1.000 &  ER HEAD & nan\\
SIEMENS & nan & MR & [0.562,0.562] & 0.640 & ZAe7b228f4 & 230051.932 & nan & nan & nan & nan & HEAD & nan\\
GE MEDICAL SYSTEMS & nan & CT & [0.449,0.449] & 625.000 & ZAbdf5826c & 145527.371 & 80.000 & 400.000 & 180.000 & 1.000 &  ER HEAD & nan\\
SIEMENS & nan & MR & [0.562,0.562] & 0.640 & ZAbdf5826c & 151328.082 & nan & nan & nan & nan & HEAD & nan\\
GE MEDICAL SYSTEMS & nan & CT & [0.449,0.449] & 625.000 & ZA977be505 & 115959.833 & 80.000 & 400.000 & 180.000 & 1.000 &  ER HEAD & 3.000\\
SIEMENS & nan & MR & [0.562,0.562] & 0.640 & ZA977be505 & 113118.462 & nan & nan & nan & nan & HEAD & nan\\
SIEMENS & nan & CT & [0.391,0.391] & 0.600 & ZA0fb7d623 & 191037.734 & 80.000 & 570.000 & 350.000 & 200.000 & HEAD & 1.500\\
SIEMENS & nan & MR & [0.562,0.562] & 0.640 & ZA0fb7d623 & 185102.800 & nan & nan & nan & nan & HEAD & nan\\
Canon Medical Systems & Aquilion ONE & CT & [0.468,0.468] & 0.500 & ZA6d523f5b & 100206.350 & 80.000 & 1000.000 & 100.000 & 100.000 & HEAD & 3.000\\
Philips Medical Systems & Ingenia CX & MR & [0.396,0.396] & 1.200 & ZA6d523f5b & 104333.810 & nan & nan & nan & nan & BRAIN & nan\\
Canon Medical Systems & Aquilion ONE & CT & [0.468,0.468] & 0.500 & ZAbde493fb & 170323.450 & 80.000 & 1000.000 & 100.000 & 100.000 & HEAD & 3.000\\
Philips Medical Systems & Ingenia CX & MR & [0.381,0.381] & 1.200 & ZAbde493fb & 172440.420 & nan & nan & nan & nan & BRAIN & nan\\
GE MEDICAL SYSTEMS & nan & CT & [0.449,0.449] & 625.000 & ZA832fd753 & 212911.089 & 80.000 & 400.000 & 180.000 & 1.000 &  ER HEAD & 3.000\\
SIEMENS & nan & MR & [0.562,0.562] & 0.640 & ZA832fd753 & 211132.547 & nan & nan & nan & nan & HEAD & nan\\
\bottomrule
\end{tabular}
}
    \caption{External dataset imaging parameters}
    \label{tab:appendix_imaging_parameters}
\end{table}

\end{document}